\documentclass[11pt]{article}
\usepackage{amsfonts}

\usepackage{amssymb}
\usepackage{latexsym}
\usepackage{graphicx}
\usepackage[english]{babel}

\begin{document}
\input epsf

% Equation numbers restart at new sections
\makeatletter
\@addtoreset{equation}{section}
\makeatother

%   \ifx\pdfoutput\undefined
 %  \usepackage[dvips]{graphicx}
  % \else
 %  \usepackage[pdftex]{graphicx}
 %  \pdfcompresslevel=9
 %  \fi
   %\usepackage{epstopdf}

   %\begin{document}
%  \usepackage{times}
%\usepackage{w-thm}
%% By default the equations are consecutively numbered. This may be changed by
%% the following command.
%% \numberwithin{equation}{section}
%%
%%
%% The usage of multiple languages is possible.
%% \usepackage{ngerman}% or
%% \usepackage[english,ngerman]{babel}
%% \usepackage[english,french]{babel}
%\begin{document}
%\input epsf
%\usepackage[]{graphicx}
%\begin{document}
%%    The information for the title page will be placed between
%%    \begin{document} and \maketitle. The order of most entries
%%    is determined by the class file and can not be changed by
%%    rearranging them. The maketitle command follows after the
%%    abstract.
%%
%%    Most of the following commands will be completed by the publisher.
%%
%%    The copyrightyear is defined in the .clo file as the first argument
%%    of the copyrightinfo command. If the copyrightyear differs from that
%%    value it might be adjusted by the following definition:
%%
%% \renewcommand{\copyrightyear}{2003}% uncomment to change the copyrightyear.
%%
 \begin{center}
{\LARGE Correlations in Hawking radiation and the infall problem}
\\
\vspace{18mm}
{\bf    Samir D. Mathur
and
Christopher J. Plumberg}
\vspace{8mm}

\vspace{8mm}

Department of Physics,\\ The Ohio State University,\\ Columbus,
OH 43210, USA\\ 
\vskip 2 mm
mathur@mps.ohio-state.edu\\
plumberg.1@buckeyemail.osu.edu
\vspace{10mm}

\end{center}

\thispagestyle{empty}

\def\p{\partial}
\def\h{{1\over 2}}
\def\be{\begin{equation}}
\def\bea{\begin{eqnarray}}
\def\ee{\end{equation}}
\def\eea{\end{eqnarray}}
\def\d{\partial}
\def\la{\lambda}
\def\eps{\epsilon}
\def\bb{\bigskip}
\def\mm{\medskip}
\newcommand{\dm}{\begin{displaymath}}
\newcommand{\edm}{\end{displaymath}}
\renewcommand{\b}{\tilde{B}}
\newcommand{\gm}{\Gamma}
\newcommand{\ac}[2]{\ensuremath{\{ #1, #2 \}}}
\renewcommand{\ell}{l}
\newcommand{\z}{\ell}
\newcommand{\newsection}[1]{\section{#1} \setcounter{equation}{0}}
\def\bb{$\bullet$}
\def\Qbar{{\bar Q}_1}
\def\QPbar{{\bar Q}_p}
\def\q{\quad}
\def\bn{B_\circ}
\def\sq{{1\over \sqrt{2}}}
\def\z{|0\rangle}
\def\o{|1\rangle}
\def\sqi{{1\over \sqrt{2}}}

\let\a=\alpha \let\b=\beta \let\g=\gamma \let\d=\delta \let\e=\epsilon
\let\c=\chi \let\th=\theta  \let\k=\kappa
\let\l=\lambda \let\m=\mu \let\n=\nu \let\x=\xi \let\r=\rho
\let\s=\sigma \let\t=\tau
\let\vp=\varphi \let\vep=\varepsilon
\let\w=\omega      \let\G=\Gamma \let\D=\Delta \let\Th=\Theta
                     \let\P=\Pi \let\S=\Sigma

\def\h{{1\over 2}}
\def\t{\tilde}
\def\r{\rightarrow}
\def\nn{\nonumber\\}
\let\bm=\bibitem
\def\Kt{{\tilde K}}
\def\b{\bigskip}

\let\p=\partial
\def\u{\uparrow}
\def\d{\downarrow}

\begin{abstract}

\b

It is sometimes believed that small quantum gravity effects can encode information as `delicate correlations' in Hawking radiation, thus saving unitarity while maintaining a semiclassical horizon. A recently derived inequality showed that this belief is incorrect: one must have order unity corrections to low energy evolution at the horizon (i.e. fuzzballs)  to remove entanglement between radiation and the hole. In this paper we take several models of `small corrections' and compute the entanglement entropy numerically; in each case this entanglement is seen to monotonically grow, in agreement with the general inequality. We also construct a model of `burning paper', where the entanglement is found to rise and then return to zero, in agreement with the general arguments of Page. We then note that the fuzzball structure of string microstates offers a version of `complementarity'. Low energy evolution is modified by order unity, resolving the information problem, while for high energy infalling modes ($E>> kT$) we may be able to replace correlators by their ensemble averaged values. Israel (and others) have suggested that this ensemble sum can be represented in the thermo-field-dynamics language as an entangled sum over two copies of the system, giving the two sides of the extended black hole diagram. Thus high energy correlators in a microstate may be approximated by correlators in a spacetime with horizons, with the ensemble sum over microstates acting like the `sewing' prescription of conformal field theory.

\end{abstract}
\vskip 1.0 true in

\newpage
\renewcommand{\theequation}{\arabic{section}.\arabic{equation}}

\def\p{\partial}
\def\r{\rightarrow}
\def\h{{1\over 2}}
\def\b{\bigskip}

\def\nn{\nonumber\\ }

\section{Introduction}
\label{intr}\setcounter{equation}{0}

Hawking's computation of radiation \cite{hawking1,hawking2} led to a serious puzzle. The radiation quanta are created in pairs, with the emitted quantum being in an entangled state with its partner which falls into the hole. By the time the hole shrinks down to planck mass, we generate a large entanglement between the emitted radiation and the hole. We are forced to either have long lived planck mass remnants, or (if the hole disappears completely) a loss of unitarity (i.e. `information loss').

Since this problem appears serious, one might wonder why there is not more effort devoted to obtaining a consensus on the issue. Relativists have of course studied this problem in quite some depth. Many studies suggested that `black holes have no hair' \cite{nohair}; if such is the case then we are forced to assume the vacuum state at the horizon that Hawking used in his emission computation, and it seems that we must go  along with his discovery of monotonically rising entanglement. Given this situation, several relativists have reconciled themselves either to the idea of information loss, or the scenario of slow leakage of information from long lived remnants. The Penrose diagrams in some treatments of loop quantum gravity also imply a kind of remnant, since the information is released not in the Hawking radiation but from a region past the nominal singularity \cite{loops}. Other treatments seek to refine our definitions of measurement in order to reconcile the density matrix obtained for the emitted quanta with the  pure state expected from a unitary evolution \cite{hartle}.

String theorists, on the other hand, have traditionally been less worried by this problem. Weak coupling computations suggest that information 
comes out in the Hawking radiation, since a manifestly unitary computation from a vibrating `effective string' reproduces exactly the greybody factory of near extremal black holes \cite{radiationall}. But at the coupling needed to make black holes one still writes the traditional metric of a hole with a smooth horizon; for example for the AdS-Schwarzschild hole one would write
\be
ds^2=-(r^2+1-{C\over r^2})dt^2+{dr^2\over r^2+1-{C\over r^2}}+r^2 d\Omega_3^2 \times d\Omega_5^2
\label{qfour}
\ee
Now one has exactly the same problem as Hawking had raised initially. So what are we supposed to do about that?

\subsection{The issue of small corrections}

A common belief has been that while Hawking did a leading order computation of the radiation, using the semiclassical metric of the hole, one should really take into account the small corrections to his computation that would result from quantum gravity effects.  These corrections could be very nonlocal and bizarre, since they could arise from tunneling processes that relate one metric to a completely different one. But one thing has always been tacitly assumed: these corrections should be `small' in the limit  ${M\over m_{pl}}\r \infty$, where $M$ is the mass of the hole and $m_{pl}$ is the planck mass. This requirement is imposed if we trust that the semi-classical black hole solution gives the leading order dynamics of the hole.  We will see below how one should encode this smallness condition mathematically in the Hawking process. Even though these corrections 
are small, the number $N$ of emitted quanta is very large
($N\sim ({M\over m_{pl}})^2$ for a 3+1 dimensional Schwarzschild hole). Thus one can hope that the smallness of the corrections can be offset by the large number of quanta, in the sense that `delicate correlations' among these $N$ quanta result in a state of the radiation that is not entangled with the hole to any significant extent by the time we reach the endpoint of evaporation. This would resolve all problems: a metric like (\ref{qfour}) would be a good semiclassical approximation, and small quantum effects would be a self consistent higher order correction that removes any information paradox at the end of the evaporation process.

But if such were the case, one has to wonder why no one produced a simple model of how such correlations could produce a unitary evolution process. That is, one could start by ignoring the details of where quantum corrections really arise from, and proceed to write down a set of `small corrections' that would lead to a pure state of radiation after the emission of $N$ quanta. If such toy models could be readily made, there would indeed be no serious paradox, since we could assume that the full quantum gravity theory would choose one of these pathways to unitarity, and the details of the exact pathway would be no more than a curiosity. 

Recently an inequality was derived which showed that there can be {\it no} such model of small corrections that will do the job of producing a pure state of the radiation \cite{mathurfuzz}. We will discuss this result in more detail below, but for now we note that this result forces us to confront the information puzzle in string theory. We cannot use `suggestive computations' that point towards a unitary radiation process to argue away the paradox; evolution in the  metric (\ref{qfour}) (with arbitrary small corrections allowed) {\it forces} us to choose between  information loss and remnants. If we wish to avoid both these possibilities, we have to find a way to modify Hawking's process by {\it order unity} at the horizon; i.e., find `hair' at the horizon. 

\subsection{The computations of this paper}

Given the above discussion, it appears important  that we understand the nature of the inequality derived in \cite{mathurfuzz}, and gain some familiarity with how it governs the growth of entanglement entropy in the black hole radiation process. 

A simple model encoding corrections to Hawking's leading order process was presented in \cite{mathurrecent}. In the leading order Hawking process the pair created at each step is not affected by pairs created at earlier steps, or by the state of the matter that initially made the hole. In the  model of \cite{mathurrecent} each created pair was allowed a small correction that depended on the emission at the {\it previous} step. This model was simple enough that it could be solved analytically. The reduction in entanglement entropy of the radiation was small, in agreement with the general inequality of \cite{mathurfuzz}. 

But since the quantum gravity effects are in general unknown, we would like to allow for arbitrarily complicated corrections; i.e., we would like these corrections to depend in any way we choose on the state of emission at earlier steps, and on the state of the initial matter that made the hole. The only restriction we would need to keep is that the overall correction to the state of each created pair be small. We are not able to solve such general models analytically, so in this paper we proceed with a numerical computation.  We look at three different models where we encode small corrections in different ways into the emission process. In each case we compute the entanglement of the emitted radiation with what is left in the hole, at different steps in the evolution process. In the leading order Hawking process (i.e. when there are no corrections) this entanglement entropy after $n$ steps of evolution is $S_{ent}=n\ln 2$. With the corrections included, we observe that we get only a slight change;  the entanglement keeps monotonically rising with $n$, and is only slightly reduced from its maximal allowed value of $n\ln 2$. Note that if the information puzzle were to be resolved by these small corrections then we would need to find that $S_{ent}$ starts {\it reducing} after a while, reaching near zero when we reach the endpoint of the evaporation process.

We then turn to an issue that has been a source of some confusion in the information puzzle: the correlations among the photons emitted from a piece of burning paper. In this case no difficulty with unitarity is expected; in line with the general analysis of \cite{page}, we expect that the entanglement of the emitted radiation with the paper at first goes up, and then goes down, reducing to zero when the paper is completely gone. We construct a model of this `burning paper', and follow the evolution of $S_{ent}$ as the evolution proceeds. We observe that $S_{ent}$ increases, reaches a maximum, and then drops back to zero, in agreement with \cite{page}.

Having the black hole models and the model of `burning paper' in front of us, we can examine why the result is so different in the two cases, and thus arrive at a basic understanding of how the two radiation processes differ. This difference was discussed in \cite{mathurfuzz}, and we revisit this issue in the context of our models, making a laboratory model that has some features of the Hawking radiation process, and noting the source of the crucial difference between this radiation process and radiation from normal warm bodies.  

Finally, we note how string theory actually gets around the information paradox. String theory black holes do have `hair';   the state at the horizon is not the vacuum state, but a very different state termed a `fuzzball'. This resolves the information paradox since we are no longer forced to the monotonic increase of entanglement created by the stretching of the vacuum at a horizon. We are now interested in understanding the `infall problem': i.e., is there a way that heavy infalling objects ($E\gg kT$) can see a universal effective behavior in all fuzzballs, perhaps modeled by the traditional geometry of the hole? Israel \cite{israel2} had noted that the two sides of the eternal Schwarzschild black hole  spacetime can be regarded as the two copies of a system used in the thermo-field-dynamics description \cite{umezawa} of a statistical ensemble. In \cite{mathurrecent} it was conjectured that in the domain $E\gg kT$ we can approximate  correlation functions by their ensemble averaged values. If we do make this replacement then we get an ensemble sum over microstates, which (following comments of Israel \cite{israel2}, Maldacena \cite{maldacena2}, Van Raamsdonk \cite{raamsdonk} and others) can be effectively replaced by the traditional black hole geometry possessing horizons. Thus we recover a version of the idea of `complementarity' \cite{thooft1,suss1}: the fuzzball construction of microstates creates the order unity corrections needed for low energy modes to carry information and resolve the information paradox, while measurements by energetic infalling objects can be approximated by computations in a spacetime with smooth horizons. 
 
\section{The inequality}
\label{ineq}\setcounter{equation}{0}

In this section we review the nature of Hawking emission, and explain the setup used to derive the inequality in \cite{mathurfuzz}. The numerical models we construct in this paper will illustrate this inequality.

\subsection{The Hawking process and entanglement entropy at the leading order}

\begin{figure}[htbp]
\begin{center}
\includegraphics[scale=.18]{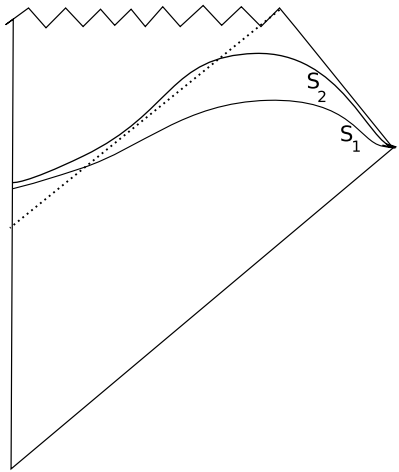}
\caption{{The Penrose diagram of a black hole formed by collapse of the `infalling matter'. The spacelike slices satisfy all the niceness conditions required for semiclassical evolution in a gravity theory.}}
\label{fthree}
\end{center}
\end{figure}

\begin{figure}[htbp]
\begin{center}
\includegraphics[scale=.18]{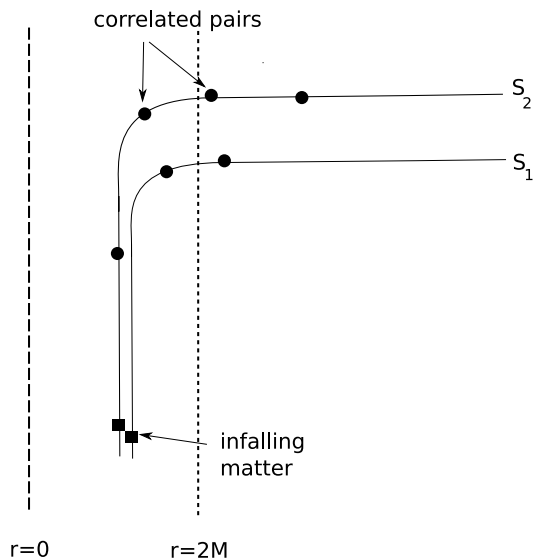}
\caption{{A schematic set of coordinates for the Schwarzschild hole. Spacelike slices are $t=const$ outside the horizon and $r=const$ inside. Infalling matter is very far from the place where pairs are created.}}
\label{ftwo}
\end{center}
\end{figure}

Consider a Schwarzschild hole with metric
\be
ds^2=-(1-{2M\over r})dt^2+{dr^2\over 1-{2M\over r}}+r^2 d\Omega_2^2
\label{qthree}
\ee
There is no singularity at the horizon, and the full spacetime made by a collapsing shell is given by the Penrose diagram shown in fig.\ref{fthree}. 
The Hawking process can be examined on the smooth slices shown in the figure, which capture the evolution up to any point where the horizon is still macroscopic,  without approaching the singularity anywhere. Since the conformal scaling of the Penrose diagram distorts the geometry of the slices, it is better to look at the slices shown in the schematic figure fig.\ref{ftwo}. The radial coordinate $r$ is plotted on the horizontal axis, and the other axis is just a way to depict evolution; it could be an Eddington-Finkelstein coordinate for instance.

Outside the hole we can make a spacelike slice by taking a $t=t_0$ surface. Inside the hole, we take a $r=r_0$ surface. These two parts of the surface can be joined by a smooth connection region ${\cal C}$.
To get evolution we must now construct a `later' slice. The $t=t_0$ part is simply changed to $t=t_0+\delta t$. Inside the hole, the forward evolution takes us to $r=r_0+\delta r$. Since we do not wish to get close to the singularity at $r=0$, we take $\delta r$ very small, such that in the entire evolution we are interested in we never get to, say, the region $r<{M\over 2}$. There is of course nothing wrong if we evolve one part of a spacelike slice forwards in time while not evolving another part of the slice; this is just the `many fingered time' of general relativity.

We now see that the $r=const$ part of this later slice has to be longer -- it has to `stretch' -- in order that it be joined to the $t=const$ part (we have used a connector segment with approximately the same intrinsic geometry as on the first slice). The stretching is gentle and smooth, with all lengthscales and timescales associated to the stretching being of order $\sim M$. This stretching thus creates pairs of quanta on the slice, with wavelengths of order $\sim M$. Decomposing this state into quanta that emerge from the hole (labelled $b$) and quanta that stay inside (labelled $c$), we get a state that is schematically of the form $Exp[\gamma b^\dagger c^\dagger]|0\rangle_b\times |0\rangle_c$ for each step of the evolution. The explicit form of the state can be found for example in \cite{giddings} for the 2-d black hole. For our purposes we only need to note that this state is entangled between the inner and outer members of the pair, and so for simplicity we break up the evolution into a set of discrete steps (with time lapse $\Delta t \sim M$), and take a simple form of the entangled state having just two terms
\be
|\Psi\rangle_{\rm pair}=\sq \z_c\z_b+\sq\o_c\o_b
\label{pairs}
\ee
(Nothing in the argument below should be affected by this simplification, or the simplification of taking discrete timesteps.
If we had a fermionic field we would in fact have just two terms in the sum.) 

At the initial timestep we have on our spacelike slice only the shell that has passed through its horizon radius, denoted by a state $|\psi\rangle_M$. At the next time step we have, in the leading order Hawking computation, the state
\be
|\Psi\rangle= |\psi\rangle_M\otimes\Big( \sq \z_{c_1}\z_{b_1}+\sq\o_{c_1}\o_{b_1}\Big)
\label{qtwoq}
\ee
If we compute the entanglement of $b_1$ with $\{M, c_1\}$ we obtain 
\be
S_{ent}=\ln 2
\ee
At the next step of evolution the slice stretches so that the quanta $|b\rangle_1, |c\rangle_1$ move away from the middle of the `connector region' ${\cal C}$, and a new pair is pulled out of the vacuum near the center of ${\cal C}$.  The full state is
\bea
|\Psi\rangle=|\psi\rangle_M&\otimes&\Big( \sq \z_{c_1}\z_{b_1}+\sq\o_{c_1}\o_{b_1}\Big)\cr
&\otimes&\Big( \sq \z_{c_2}\z_{b_2}+\sq\o_{c_2}\o_{b_2}\Big)
\label{qtwoq2}
\eea
If we compute the entanglement of the set $\{b_1, b_2\}$ with $\{M, c_1, c_2\}$, we find
\be
S_{ent}=2\ln 2
\label{ent2}
\ee
Continuing this process, after $N$ steps we get, in the leading order Hawking computation,
\bea
|\Psi\rangle= |\psi\rangle_M&\otimes&\Big( \sq \z_{c_1}\z_{b_1}+\sq\o_{c_1}\o_{b_1}\Big)\cr
&\otimes&\Big( \sq \z_{c_2}\z_{b_2}+\sq\o_{c_2}\o_{b_2}\Big)\cr
&\dots&\cr
&\otimes&\Big( \sq \z_{c_N}\z_{b_N}+\sq\o_{c_N}\o_{b_N}\Big)
\label{qtwoq3}
\eea
The entanglement entropy of the  $\{ b_i\}$ with $ \{ M,  c_i\}$ is
\be
S_{ent}=N\ln 2
\label{ent}
\ee
Since this entanglement keeps growing with $N$, we get the Hawking problem mentioned above.

\subsection{Allowing small corrections}

Let the state at time step $t_n$ be written as  $|\Psi_{M, c, b}(t_n)\rangle$. Here $\{M\}$ denotes the  matter shell that fell in to make the black hole, $\{c\}$ denotes the quanta that have been created at earlier steps in the evolution and $\{b\}$ denotes the set of all $b$ quanta that have been created in all earlier steps. The state $\Psi$ is entangled between the $\{M,c\}$ and $\{b\}$ parts; it is not a product state. We assume nothing about its detailed structure. 

We can choose a basis of orthonormal states $\psi_n$ for the $\{M,c\}$ quanta inside the hole, and an orthonormal basis $\chi_n$ for the quanta $\{b \}$ outside the hole, such that 
\be
|\Psi_{M, c, b}(t_n)\rangle=\sum_{m,n} C_{mn} \psi_m\chi_n
\ee
It is  convenient to make unitary transformations on the $\psi_m, \chi_n$  so that we get
\be
|\Psi_{M, c, b}(t_n)\rangle=\sum_{i} C_{i} \psi_i\chi_i
\label{stateone}
\ee
We compute the reduced density matrix describing the $b_i$ quanta outside the hole
\be
\rho_{ij}=|C_i|^2 \delta_{ij}
\ee
The entanglement entropy at time $t_n$ is then
\be
S_{ent}(t_n)=-\sum_i |C_i|^2 \ln |C_i|^2
\label{entone}
\ee
In the leading order  evolution we would have  at time step $t_{n+1}$:
\be
|\Psi_{M, c, b}(t_n)\rangle\r |\Psi_{M, c, b}(t_n)\rangle\otimes \Big [\sqi |0\rangle_{c_{n+1}}|0\rangle_{b_{n+1}}+\sqi |1\rangle_{c_{n+1}}|1\rangle_{b_{n+1}}\Big ]
\label{leading}
\ee
where the term in box brackets denotes the state of the newly created pair. 

Let us now write down the modifications  to this evolution from timestep $t_n$ to timestep $t_{n+1}$ that will encode  the small corrections we wish to allow:

\b

(A) For the quanta that have left the hole at earlier timesteps, we have no change
\be
\chi_i\r \chi_i
\label{chifixed}
\ee
This is the case even for the example of burning paper: quanta that have left the paper and been collected outside do not participate in the next step of the evolution of the paper.

\b

(B) In the leading order evolution there was no change to the state of $\{ M,c\}$ inside the hole, since the part of the spacelike slice carrying these quanta did not evolve forwards in time. We will now allow a completely general unitary evolution of this state to any other state formed by these quanta. 

\b

(C) In the leading order Hawking computation the state of the newly created pair was $|\Psi\rangle_{\rm pair}$ (eq. (\ref{pairs})). We now allow the state of this pair to lie in a 2-dimensional subspace, spanned by $|\Psi\rangle_{\rm pair}\equiv S^{(1)}$ and an orthogonal vector $S^{(2)}$:
\bea
S^{(1)}&=&\sqi  |0\rangle_{c_{n+1}}|0\rangle_{b_{n+1}}+\sqi  |1\rangle_{c_{n+1}}|1\rangle_{b_{n+1}}\cr
S^{(2)}&=&\sqi  |0\rangle_{c_{n+1}}|0\rangle_{b_{n+1}}-\sqi  |1\rangle_{c_{n+1}}|1\rangle_{b_{n+1}}\cr
&&
\label{set}
\eea

\b

Putting together (B) and (C), we see that the most general evolution allowed for the state in the hole is 
\be
\psi_i\r \psi_i^{(1)}S^{(1)}+\psi_i^{(2)}S^{(2)}
\ee
where $\psi_i^{(1)}, \psi_i^{(2)}$ are any two states of $\{M,c\}$ (the initial matter and the infalling members of Hawking pairs produced at earlier steps). As promised in (B),  these states be arbitrary, but note that unitarity of evolution gives
\be
|| \psi_i^{(1)}||^2+|| \psi_i^{(2)}||^2=1
\ee
We thus get the evolution
\be
|\Psi_{M, c, b}(t_{n+1})\rangle=\sum_{i} C_{i} [\psi_i^{(1)}S^{(1)}+\psi_i^{(2)}S^{(2)}]~ \chi_i
\label{statetwo}
\ee

We can write the state (\ref{statetwo}) as
\be
|\Psi_{M, c, b}(t_{n+1})\rangle=\Big [ \sum_{i} C_{i} \psi_i^{(1)} \chi_i\Big ]S^{(1)} + \Big [ \sum_{i} C_{i} \psi_i^{(2)} \chi_i\Big ]S^{(2)} \equiv  \Lambda^{(1)}S^{(1)}+  \Lambda^{(2)}S^{(2)}
\label{state}
\ee
Here we have defined the states
\be
\Lambda^{(1)}= \sum_{i} C_{i} \psi_i^{(1)} \chi_i, ~~\Lambda^{(2)}= \sum_{i} C_{i} \psi_i^{(2)} \chi_i
\ee
Since $S^{(1)}, S^{(2)}$ are orthonormal, normalization of $|\Psi_{M, c, b}(t_{n+1})\rangle$ implies that
\be
||\Lambda^{(1)}||^2+||\Lambda^{(2)}||^2=1
\ee

We can now state precisely what it means for the quantum corrections to be small. If the evolution from step $n$ to step $n+1$ is to be close to the semiclassical one in any sense then we must have produced mostly the state $|\Psi\rangle_{\rm pair}\equiv S^{(1)}$ and only a small amount if $S^{(2)}$. Thus we write
\be
|| \Lambda^{(2)}||<\epsilon, ~~~\epsilon\ll 1
\label{cond1}
\ee
If there is no such bound, then we will say that the corrections to the Hawking evolution are `order unity'. 

The result of \cite{mathurfuzz} says that with the smallness condition (\ref{cond1}) we get a minimal increase of entanglement at each step
\be
S_{n+1}-S_n>\ln 2-2\epsilon
\label{result}
\ee

It is important to note that the smallness condition (\ref{cond1}) is placed on the evolution inside a spacetime region that is size $\sim M$ on each side. Thus we are just writing down in mathematical terms the requirement that the full state and the semiclassical state differ by a small amount in the local process of evolution near the horizon by one timestep of order $M$ (the timescale required to create one Hawking pair). We are {\it not} requiring that the full state of all the $\{b\}$ quanta be close to the one that we would obtain by semiclassical evolution. For a state made of many quanta, or in the process of evolution by many timesteps, we can get large changes from the cumulative effects of small changes. For example suppose we have a product state $|\Psi\rangle=\prod_{i=1}^k \psi_i$, and another such state $|\Psi'\rangle=\prod_{i=1}^k\psi'_i$. Even if each $|\psi\rangle$ is close to the corresponding $|\psi'\rangle$ (i.e., $\langle\psi|\psi'\rangle=1-\epsilon$), we get a small overlap between the total wavefunctions
\be
\langle \Psi|\Psi'\rangle=(1-\epsilon)^k\r 0 ~~{\rm when}~~k\r \infty
\ee
By contrast, in our argument above we are only requiring that in a spacetime ball of size $\sim M$ the full evolution and the semiclassical one differ by a small amount, something that we {\it must} have if we are to say that the horizon evolution is close to the semiclassical one.

\section{Numerical models}
\label{secthree}\setcounter{equation}{0}

In this section we make several models where we add small corrections to the leading order Hawking process, and numerically compute the entanglement entropy as the evolution proceeds. We will deliberately choose quite complicated evolution models, so that our conclusions are not governed by some residual symmetry in our choice of evolution.   In each case we find that the entanglement keeps growing, in accordance with the general theorem above.

\subsection{Model A}

We will describe the structure of this model in some detail, so that we can point out all the constraints we have to be careful to incorporate in any such model. 

\subsubsection{Evolution in the model}

In the leading order Hawking process the state of the newly created pair is given by $S^{(1)}$ (eq.(\ref{set})). This state does not depend on the matter $M$ which made the hole or the negative energy quanta $\{ c\}$ that have been created at previous steps (which are also going to be in the hole). For our first model, we will let the small corrections be independent of $M$, but will allow a dependence on the $\{ c\}$. (In the model that follows we will include dependence on $M$ as well.) The case where we get a correction only from the $c$ quantum created at the previous step was solved analytically in \cite{mathurrecent}. In our present numerical model we will allow a very  `nonlocal' dependence on the $\{ c \}$; this means that the emission at step $n+1$ is affected not only by the $c$ quanta created in the last few steps, but by the $c$ quanta created at all previous steps. 

From (\ref{state}) we see that to define the evolution with small corrections we need to find the state at timestep $n+1$
\be
|\Psi_{M, c, b}(t_{n+1})\rangle\equiv |\Psi\rangle_{n+1}= \Lambda^{(1)}_nS^{(1)}_{n+1}+  \Lambda^{(2)}_nS^{(2)}_{n+1}
\label{basic1}
\ee
where $S^{(1)}, S^{(2)}$ are given in (\ref{set}). Thus our goal is to define $\Lambda^{(1)}, \Lambda^{(2)}$ for the model. We proceed in the following steps.

\bigskip

(i) At step $n=0$ we have just the matter $M$ on our spacelike slice. There are no $c,b$ quanta.

\b

(ii) At timestep $n=1$ the state is the leading order Hawking state
\be
|\Psi\rangle_{n=1}=\sq \z_{c_1}\z_{b_1}+\sq\o_{c_1}\o_{b_1}
\label{firststep}
\ee
since we are not taking into account any dependence in $M$, and there are no $c$ quanta from earlier steps.
 
\b
 
(iii) At any step $n$, we write the complete state (\ref{state}) in the following basis. Consider the states of the $\{ b \}$ quanta. There are $n$ steps in the emission process upto this point, and at each step we either have no emission ($|0\rangle_{b_i}$) or an emitted quantum ($|1\rangle_{b_i}$). Thus we get a basis of orthonormal states of the $\{  b \}$ quanta by taking sets of $n$ numbers, with each number being a $0$ or a $1$. Thus there are $2^n$ possible states of the $\{ b \}$. We call these states $\chi^{(b)}_k$, labelled by an index $k=1, \dots 2^n$: 
We write
\bea
\chi^{(b)}_1&=&00\dots 00 \nn\cr
\chi^{(b)}_2&=&00\dots 01\nn\cr
&\dots&\nn\cr
\chi^{(b)}_{2^n}&=&11\dots 11
\label{tenq}
\eea
A similar basis  $\psi^{(c)}_k$ exists for the $\{ c\}$ quanta. 
Thus we can expand the state at timestep $n$ as
\be
|\Psi\rangle_n=\sum_{k,l} C_{kl} \psi^{(c)}_k\chi^{(b)}_l
\label{stepn}
\ee
with
\be
\sum_{k=1}^{2^n}\sum_{l=1}^{2^n}|C_{kl}|^2=1
\label{norm1}
\ee

\b

(iv) The $\{b\}$ quanta emitted at timesteps $1, \dots n$ cannot be altered by evolution at step $n+1$, since they have already left the system and we can imagine they have been collected outside. Thus
\be
\chi_l^{(b)}\r \chi_l^{(b)}
\label{chi1}
\ee

\b

(v) We need to choose some way to incorporate the dependence of the small correction on the $\{ c\}$ quanta. We choose a small parameter $\epsilon$ that will govern the smallness of the corrections. 

Now we wish to make a definite but quite arbitrarily chosen rule to govern the amount of correction to apply to a state containing a given component $\psi^{(c)}_k$ for the $\{ c\}$ quanta. Take a state $\psi_k^{(c)}$; let it be for example
\be
1001110
\label{exstate}
\ee
Thus we have chosen $n=7$, and in this part of the amplitude we have  emission at step $1$, no emission at steps $2,3$, then emission at steps $4,5,6$, and no emission at step $7$. Let $n_1^k$ be the count of {\it odd numbered} steps where we have a $1$, and let $n_2^k$ be the count of {\it even numbered} steps where we have a $0$. Finally, let
\be
n_3^k=n_1^k+n_2^k
\ee
Thus in the above example we find a $1$ at steps $1$ and $5$, so $n_1^k=2$. We find a $0$ at step $2$, so we get $n_2^k=1$. Finally, $n_3^k=n_1^k+n_2^k=3$.  

The size of the small correction on the part of the wavefunction $\psi_k^{(c)}$ will be governed by
\be
{n_3^k\over n}\epsilon\le \epsilon
\label{small1}
\ee

\b

(vi) In the leading order Hawking evolution the state of the $\{ c\}$ quanta produced at earlier steps is not altered when the new pair is created. But in our model with corrections we will allow an arbitrary change to the state of these quanta. Let us first define our construction of the state $\Lambda^{(2)}$. For the part of the wavefunction with the $\{ c\}$ quanta in the state $\psi_k^{(c)}$ we take the evolution
\be
\psi_k^{(c)} \r \psi_k^{(c),{\rm cl~ rot}}
\ee
where the superscript `${\rm cl ~rot}$' means that we cycle the entries in $\psi_k^{(c)}$ in a clockwise rotation. Thus for the state (\ref{exstate}) we have
\be
1001110\r 0100111
\ee
We can see that this is a very `nonlocal' effect on the $c$ quanta. With this definition, we have
\be
\Lambda^{(2)}_n=\sum_{k=1}^{2^n}\sum_{l=1}^{2^n} C_{kl} \Big ({n_3^k\over n} \epsilon\Big ) \psi_k^{(c),{\rm cl~ rot}}\chi_l^{(b)}
\label{lambda2}
\ee

\b

(vii) In a similar way we define $\Lambda^{(1)}$. This time we can take a different modification to $\psi_k^{(c)}$, and we choose an {\it anti}-clockwise rotation
\be
\psi_k^{(c)}\r \psi_k^{(c),{\rm acl~rot}}
\ee
Thus for the state (\ref{exstate}) we would get
\be
1001110\r 0011101
\ee
We set
\be
\Lambda^{(1)}_n=\sum_{k=1}^{2^n}\sum_{l=1}^{2^n} C_{kl} \sqrt{\Big (1-({n_3^k\over n} \epsilon)^2\Big )} \psi_k^{(c),{\rm acl~ rot}}\chi_l^{(b)}
\label{lambda1}
\ee
where we have chosen the factor $\sqrt{\Big (1-({n_3^k\over n} \epsilon)^2\Big )}$ to get the correct normalization of the overall state, something we will check shortly.

\b

(viii) To summarize, if the state at timestep $n$ is (\ref{stepn}), then
the  state at step $n+1$ is given by
\be
|\Psi\rangle_{n+1}=\Lambda^{(1)}_n\Big (\sq \z_{c_{n+1}}\z_{b_{n+1}}+\sq \o_{c_{n+1}}\o_{b_{n+1}}\Big )+\Lambda^{(2)}_n\Big (\sq \z_{c_{n+1}}\z_{b_{n+1}}-\sq \o_{c_{n+1}}\o_{b_{n+1}}\Big )
\ee
where $\Lambda^{(1)}$ is given by (\ref{lambda1}) and $\Lambda^{(2)}$ is given by (\ref{lambda2}). This state $|\Psi\rangle_{n+1}$ has the following features. The state of the quanta $c_{n+1}, b_{n+1}$ is close to the leading order Hawking state $S^{(1)}$, since $\Lambda^{(2)}_n$ is proportional to the smallness parameter $\epsilon$. Apart from this requirement of smallness, we have been quite general in our choice of how the small perturbations are generated: they can depend in quite a nontrivial way on all the $c$ quanta created at earlier steps, and the state of these $c$ quanta in the hole has also been allowed to change in a nontrivial way at the evolution timestep $n+1$.

\subsubsection{Checking unitarity of evolution and smallness of the perturbation}\label{section9}

In our model of evolution above we constructed a state $|\Psi\rangle_{n+1}$ for each possible state $|\Psi\rangle_n$. We would now like to check that this evolution we defined was unitary; this is of course a necessary condition for it to be a valid evolution rule. The second thing we would like to check is that the evolution is close to the leading order Hawking evolution as far as the state of the newly created pair is concerned; this requirement comes from the assumption that the black hole has  a `traditional horizon' where the matter state is close to the vacuum in a `good slicing'. 

\b

(a) A complete orthonormal basis of states at timestep $n$ is given by the $\psi^{(c)}_k \chi^{(b)}_l$, where $\psi_k^{(c)}$ and $\chi^{(b)}_l$ are complete orthonormal bases for the $\{ c\}$ and $\{ b \}$ quanta respectively.  We have $\chi^{(b)}_l\r \chi^{(b)}_l$, so all we need to ensure unitarity is that the $\psi_k^{(c)}$ evolve to a set of orthonormal states. We have
\be
\psi_k^{(c)}\r \psi_k^{'(c)}
\ee
\be
\psi_k^{'(c)}= \sqrt{\Big (1-({n_3^k\over n} \epsilon)^2\Big )} \psi_k^{(c),{\rm acl~ rot}}
S^{(1)}_{n+1}+\Big ({n_3^k\over n} \epsilon\Big ) \psi_k^{(c),{\rm cl~ rot}}  S^{(2)}_{n+1}
\ee
where
\be
S^{(1)}_{n+1}=\Big (\sq \z_{c_{n+1}}\z_{b_{n+1}}+\sq \o_{c_{n+1}}\o_{b_{n+1}}\Big ), ~~~S^{(2)}_{n+1}=\Big (\sq \z_{c_{n+1}}\z_{b_{n+1}}-\sq \o_{c_{n+1}}\o_{b_{n+1}}\Big )
\ee
Note that
\be
\langle S^{(1)}_{n+1}|S^{(1)}_{n+1}\rangle=1, ~~~\langle S^{(2)}_{n+1}|S^{(2)}_{n+1}\rangle=1, ~~~\langle S^{(1)}_{n+1}|S^{(2)}_{n+1}\rangle=0
\ee
We find
\bea
\langle \psi_k^{'(c)}|\psi_{k'}^{'(c)}\rangle &=&\sqrt{\Big (1-({n_3^k\over n} \epsilon)^2\Big )}\sqrt{\Big (1-({n_3^{k'}\over n} \epsilon)^2\Big )}\langle  \psi_k^{(c),{\rm acl~ rot}}| \psi_{k'}^{(c),{\rm acl~ rot}}\rangle\nn
&&+~~\Big ({n_3^k\over n} \epsilon\Big )\Big ({n_3^{k'}\over n} \epsilon\Big )\langle  \psi_k^{(c),{\rm cl~ rot}}| \psi_{k'}^{(c),{\rm cl~ rot}}\rangle\nn
&=&\delta_{kk'} \Big [\Big (1-({n_3^k\over n} \epsilon)^2\Big )+\Big ({n_3^k\over n} \epsilon\Big )^2\Big ]\nn
&=&\delta_{kk'}
\eea
where in the second step we have used the fact that clockwise and anticlockwise rotations are operations that map the orthonormal set $\psi^{(c)}_k$ to an orthonormal set of states:
\be
\langle  \psi_k^{(c),{\rm acl~ rot}}| \psi_{k'}^{(c),{\rm acl~ rot}}\rangle=\delta_{kk'}, ~~~\langle  \psi_k^{(c),{\rm cl~ rot}}| \psi_{k'}^{(c),{\rm cl~ rot}}\rangle=\delta_{kk'}
 \ee
 Thus we see that out evolution from timestep $n$ to timestep $n+1$ is indeed a unitary operation. 
 
 \b

(b) We now wish to verify the smallness condition (\ref{cond1}). The state $\Lambda^{(2)}_n$ is given in eq.(\ref{lambda2}). Noting the orthonormality of our basis vectors
\be
\langle \chi^{(b)}_l|\chi^{(b)}_{l'}\rangle=\delta_{ll'}, ~~~~\langle \psi^{(c)}_k|\psi^{(c)}_{k'}\rangle=\delta_{kk'}
\ee
we find
\bea
||\Lambda^{(2)}_n||^2&=&\sum_{k=1}^{2^n}\sum_{l=1}^{2^n} |C_{kl}|^2 \Big ({n_3^k\over n} \epsilon\Big )^2\nn
&\le & \epsilon^2\sum_{k=1}^{2^n}\sum_{l=1}^{2^n} |C_{kl}|^2  \nn
&=&\epsilon^2
\eea 
where in the first step we have used (\ref{small1}) and in the second step we have used (\ref{norm1}). Thus we have
\be
||\Lambda_n^{(2)}||\le \epsilon
\ee
and we ensure smallness of our corrections by taking $\epsilon\ll 1$. 

\subsubsection{The entanglement entropy: numerical results}

\begin{figure}[htbp]
\begin{center}
\includegraphics[scale=.58]{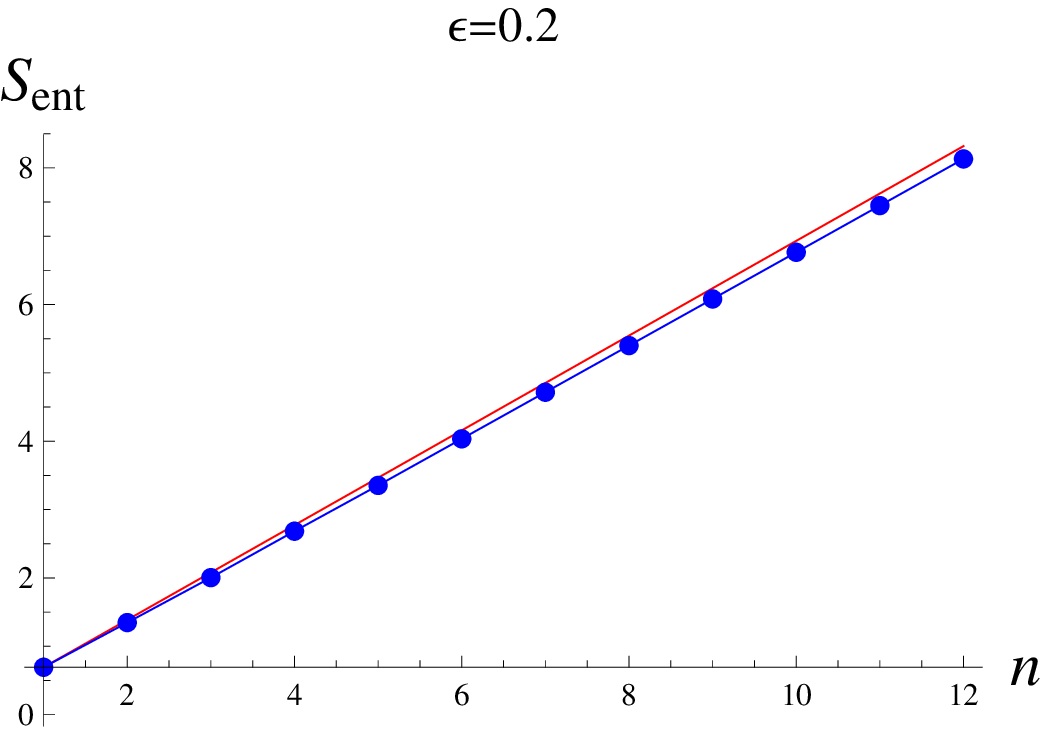}\hskip 1 true in
\includegraphics[scale=.58]{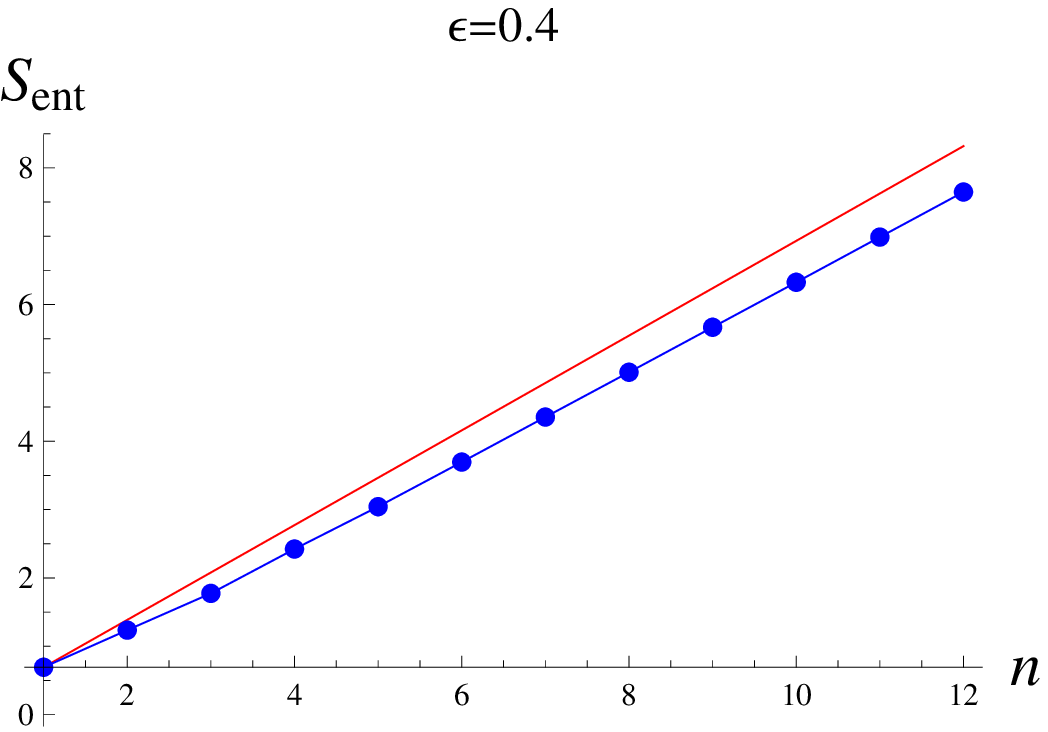}
\caption{{Entanglement entropy as a function of timestep in model A, for  two different values of the smallness parameter $\epsilon$.}}
\label{fn4m}
\end{center}
\end{figure}

At any timestep $n$ we have an entangled state of the $\{c\}$ quanta and the $\{ b\} $ quanta. We trace over the states of the $\{ c\}$ quanta, obtaining a reduced density matrix $\rho^{(b)}_n$ describing the $\{ b \}$ quanta. We then compute the entanglement entropy
\be
S_{ent}(n)=-Tr \rho^{(b)}_n\ln \rho^{(b)}_n
\ee
The computation of $\rho^{(b)}_n$ and of $S_{ent}(n)$ are done numerically. The value for $S_{ent}(n)$ is plotted as a function of $n$ in fig.\ref{fn4m}, for two different values of the smallness parameter $\epsilon$ ($\epsilon=0.2$ and $\epsilon=0.4$).   The upper red line in each case shows the value of $S_{ent}(n)$ that we would obtain for $\epsilon=0$; this is just the graph
\be
S_{ent}^{\epsilon=0}(n)=n\ln 2
\ee
and gives the entanglement entropy that we would get in the leading order Hawking computation. 

From the graphs in fig.\ref{fn4m} we see that the entanglement entropy $S_{ent}$ is only slightly reduced below its leading order value, and in particular, it keeps on monotonically rising with $n$. This monotonic rise is the essential point of the  result established in \cite{mathurfuzz}, and is the point that we wished to illustrate with our numerical models of Hawking radiation.

\b

\subsection{Model B}

In Model A above we did not incorporate any dependence of the corrections on the matter $M$ which initially made the hole. In Model B we will incorporate such a correction, but keep other features the same as those in Model A.

\begin{figure}[htbp]
\begin{center}
\includegraphics[scale=.58]{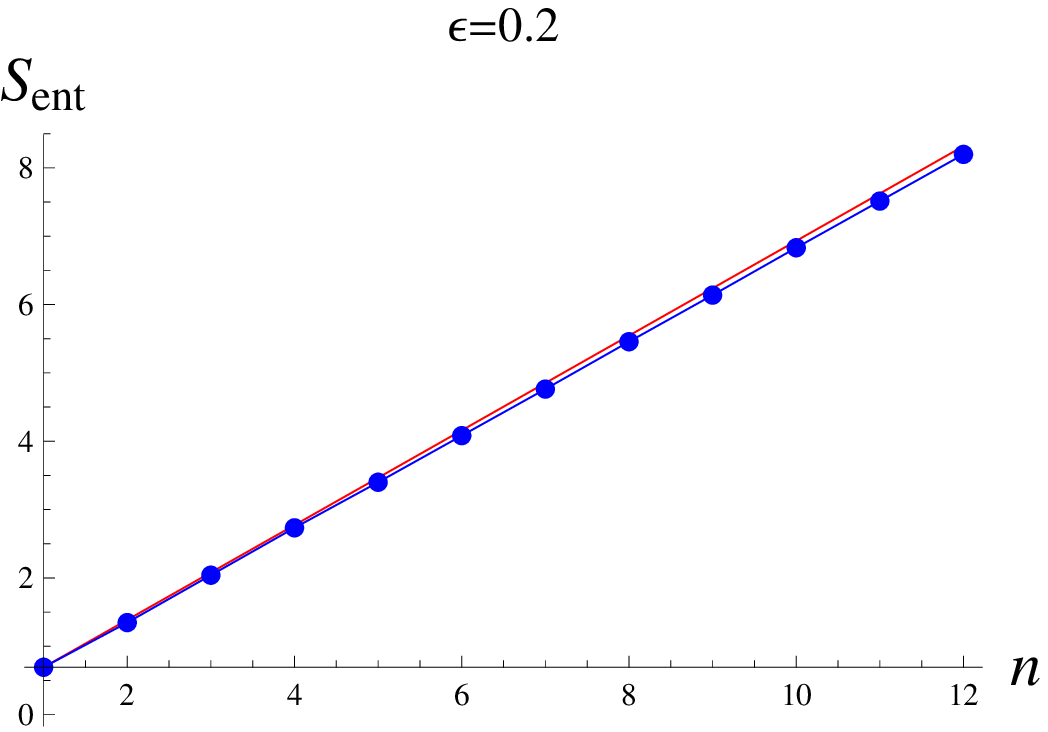}\hskip 1 true in
\includegraphics[scale=.58]{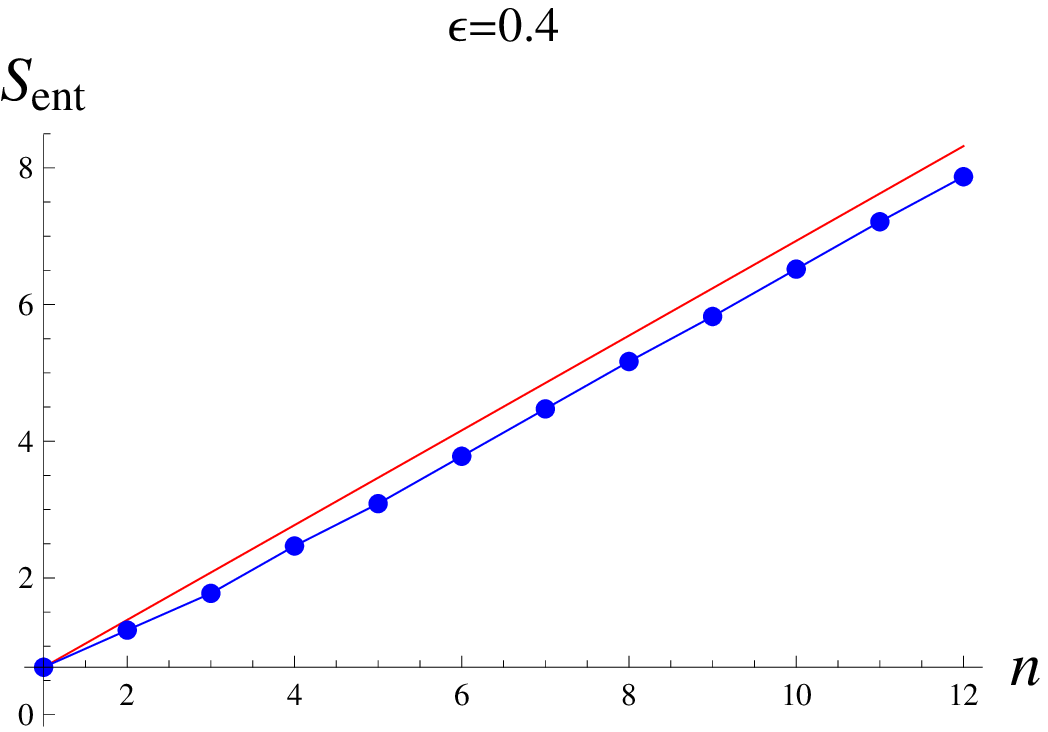}
\caption{{Entanglement entropy as a function of timestep in model B, for  two different  values of the smallness parameter $\epsilon$. In the first case the matter is given by $M= \{100111010110\}$ and in the second by $M=\{ 110100010010\}$.}}
\label{fn5m}
\end{center}
\end{figure}

We will evolve the model for $n=12$ timesteps, so we let the matter be described by a set of $12$ `bits'. For example we may take
\be
M=100111010110
\label{ex2}
\ee
We now have the following evolution rules. Timestep $n=0$ has just the matter $M$. Timestep $n=1$ has the same form (\ref{firststep}) as in model $A$. 
In evolving from timestep $n=1$ to timestep $n=2$, we look at the first element of $M$. In the example (\ref{ex2}) this is a $1$. We therefore keep the same evolution rule as in model A. For evolution from timestep $n=2$ to $n=3$,  we look at the second entry of $M$. In the example we have taken this entry is $0$. Thus for this step of evolution we set the value of $\epsilon$ to zero; i.e., 
\be
\Lambda^{(2)}_2=0, ~~~\Lambda^{(1)}=\sum_{k=1}^{2^n}\sum_{l=1}^{2^n} C_{kl}  \psi_k^{(c),{\rm acl~ rot}}\chi_l^{(b)}
\label{lambda1p}
\ee
We continue in this way, keeping the evolution rules of Model A when we have a $1$ in the corresponding entry in $M$, and changing this rule by setting $\epsilon=0$ when this entry is $0$. Thus the overall state generated has a dependence on the choice of $M$. The unitarity and smallness conditions follow just as in model A. We define the density matrix $\rho^{(b)}$ and $S_{ent}(n)$ as before. In fig.\ref{fn5m} the first plot shows $S_{ent}$ for $\epsilon=0.2$ and the matter in the state $M= \{100111010110\}$. The second plot gives $S_{ent}$ for $\epsilon=0.4$ and $M=\{ 110100010010\}$.  Again, we see that there is very little correction to the leading order Hawking result, and in particular 
$S_{ent}$ keeps rising monotonically.

\subsection{Model C}

In the general proof of \cite{mathurfuzz} that the entanglement entropy keeps monotonically increasing, we assumed that the state of the created pair lay in the subspace spanned by $S^{(1)}$ and $S^{(2)}$ (eq.(\ref{set})). Our models above have kept to this choice. But this choice of a 2-dimensional subspace was a simplification which helped us illustrate the essence of the problem; in general we can approximate the situation at each timestep by using a finite dimensional subspace, with the accuracy of the approximation improving with the dimension of the chosen subspace. 
While the essential result that the entanglement keeps rising is not expected to be affected by the exact choice of chosen subspace, it would be helpful to see this fact in an explicit model. Thus we consider in our third model (Model C) a situation were we have a {\it four} dimensional subspace in place of the two-dimensional one. 

\subsubsection{Evolution in the model}

The evolution steps are described as follows:

\b

(i) We take no dependence on the initial matter $M$ which made the hole. At timestep $n=1$ we generate a $c,b$ pair in the state $S^{(1)}$
\be
|\Psi\rangle_{n=1}=\sq \z_{c_1}\z_{b_1}+\sq\o_{c_1}\o_{b_1}
\ee

\b

(ii) Suppose we have evolved to timestep $n$. At timestep $n+1$ we will create a new pair in a state that is a linear combination of the following four states
\bea
S^{(1)}_{n+1}&=&\sq \z_{c_{n+1}}\z_{b_{n+1}}+\sq \o_{c_{n+1}}\o_{b_{n+1}}\nn
S^{(2)}_{n+1}&=&\sq \z_{c_{n+1}}\z_{b_{n+1}}-\sq \o_{c_{n+1}}\o_{b_{n+1}}\nn
S^{(3)}_{n+1}&=&\z_{c_{n+1}}\o_{b_{n+1}}\nn
S^{(4)}_{n+1}&=&\o_{c_{n+1}}\z_{b_{n+1}}
\eea
Similar to the evolution (\ref{basic1}), we now have
\be
|\Psi\rangle_{n+1}=\Lambda^{(1)}_nS^{(1)}_{n+1}+  \Lambda^{(2)}_nS^{(2)}_{n+1}
+\Lambda^{(3)}_nS^{(3)}_{n+1}+  \Lambda^{(4)}_nS^{(4)}_{n+1}
\label{full1}
\ee
As always, the state of new pair will be `mostly' $S^{(1)}$, with a small correction involving the other three states.

\b

(iii) Let the state at timestep $n$ be expanded as (\ref{stepn}). We still have (\ref{chi1}), but we take a different rule for the evolution of the $\psi_k^{(c)}$, which is a set of $1$ and $0$ entries in a list of length $n$. Let these entries be called $w^k_j, j=1,2,\dots n$; thus $w^k_j$ takes the values $0$ or $1$ for each $j$. We wish to have a stronger dependence on $w^k_n$, the state of the $c$ quantum that has just been emitted, than on $w^k_{n-1}$, the quantum emitted before that, and so on.  Thus we define
\bea
w^{k,(2)} &=&\sum_{j=1 }^n {1\over 2^{n-j+1}} w^k_j\nn
w^{k,(3)} &=&\sum_{j=1, \, j \, odd}^n {1\over 2^{n-j+1}} w^k_j\nn
w^{k,(4)} &=&\sum_{j=1,\,  j\, even}^n {1\over 2^{n-j+1}} w^k_j\nn
\eea
which all depend on  the $w^k_j$ but more on the `recent' $w^k_j$ entries. We have 
\be
w^{k, (a)}\le 1, ~~~a=2,3,4
\ee
We choose a parameter $\epsilon\ll 1$, and will let the states $S^{(2)}, S^{(3)}, S^{(4)}$ appear in the perturbed pair state with amplitudes proportional to $\epsilon w^{k,(2)}, \epsilon w^{k,(3)},\epsilon w^{k,(4)}$ respectively.

\b

(iv) In the models above we had deformed the state of the $\{ c\}$ quanta in the hole by a clockwise or an anticlockwise rotation of the quanta. For our present model we change this deformation to a `bit swap' where we replace all entries $1$ in $\psi_k^{(c)}$  with $0$ and all 
entries $0$ with $1$:
\be
\psi_k^{(c)}\r \psi_k^{'(c)}=\psi_k^{(c), {\rm bit\, swap}}
\ee

\b

(v) We now define, analogous to (\ref{lambda2})
\bea
\Lambda^{(2)}_n&=&\sum_{k=1}^{2^n}\sum_{l=1}^{2^n} C_{kl} \Big (\epsilon w^{k, (2)}\Big ) \psi_k^{(c),{\rm bit\, swap}}\chi_l^{(b)}\nn
\Lambda^{(3)}_n&=&\sum_{k=1}^{2^n}\sum_{l=1}^{2^n} C_{kl} \Big (\epsilon w^{k, (3)}\Big ) \psi_k^{(c),{\rm bit\, swap}}\chi_l^{(b)}\nn
\Lambda^{(4)}_n&=&\sum_{k=1}^{2^n}\sum_{l=1}^{2^n} C_{kl} \Big (\epsilon w^{k, (4)}\Big ) \psi_k^{(c),{\rm bit\, swap}}\chi_l^{(b)}
\eea
and finally, analogous to (\ref{lambda1})
\bea
\Lambda^{(1)}_n&=&\sum_{k=1}^{2^n}\sum_{l=1}^{2^n} C_{kl} \sqrt{\Big (1-\epsilon^2 [(w^{k, (2)})^2+(w^{k, (3)})^2+(w^{k, (4)})^2]\Big )} \psi_k^{(c),{\rm bit\, swap}}\chi_l^{(b)}
\eea
where the term under the square root is chosen to make the evolution unitary. Substituting these $\Lambda_n^{(i)}, i=1, \dots 4$ in (\ref{full1}), we get our full evolution rule, for which we can check unitarity and the smallness requirement just as we did for Model A in section \ref{section9}. 

\begin{figure}[htbp]
\begin{center}
\includegraphics[scale=.58]{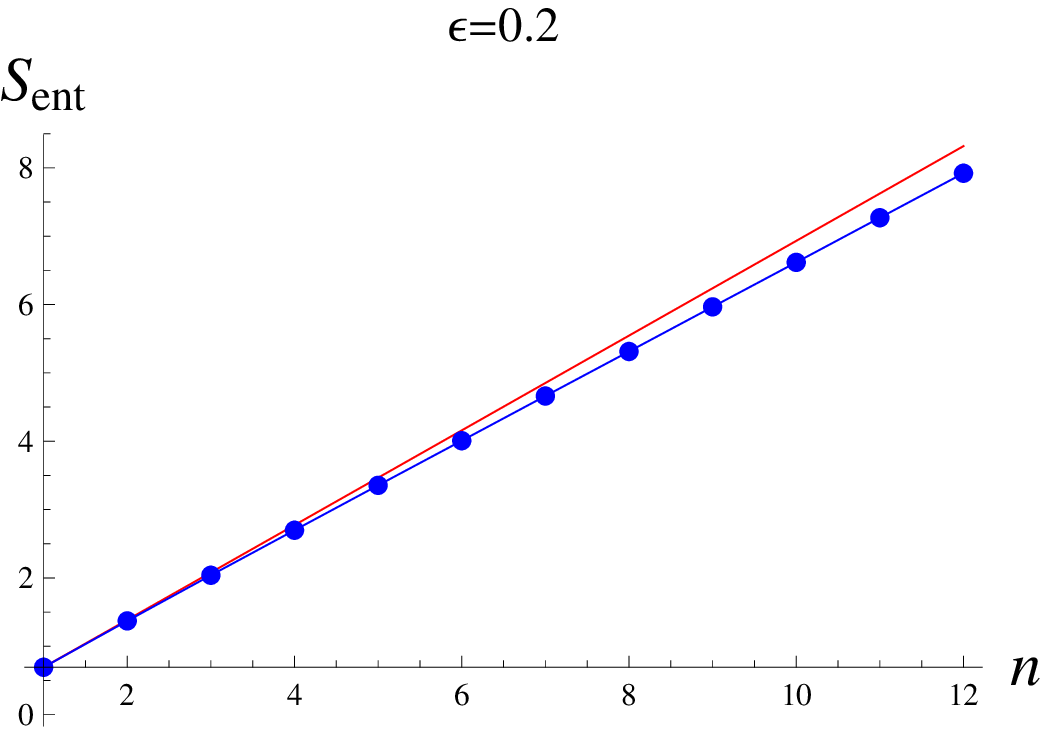}\hskip 1 true in
\includegraphics[scale=.58]{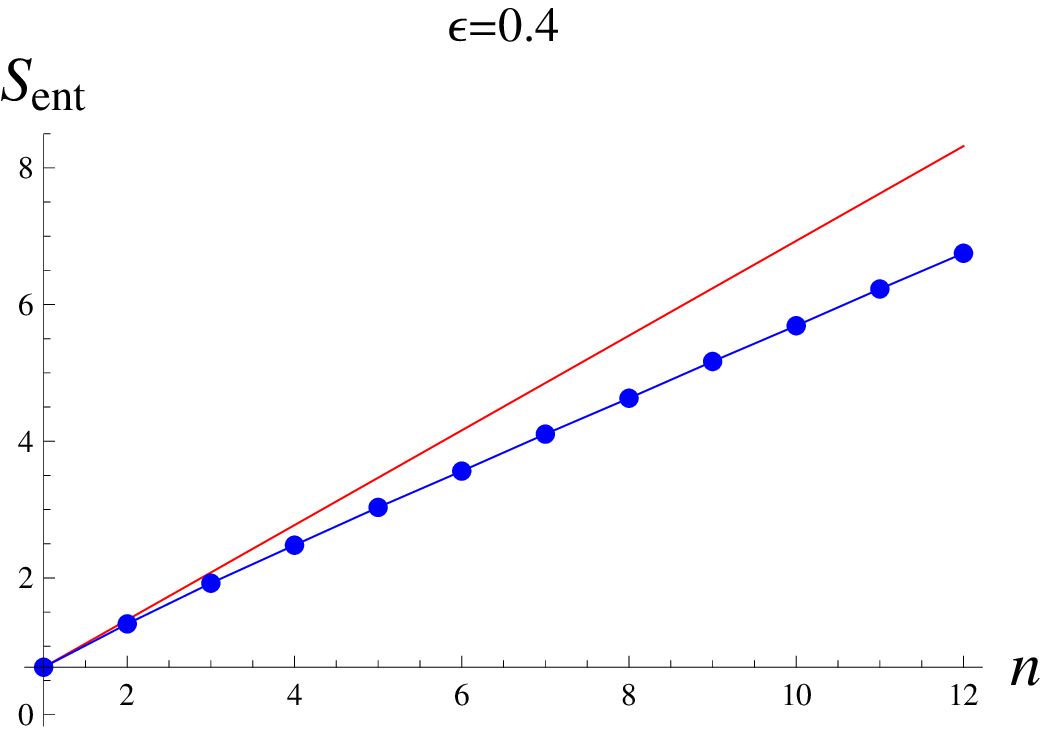}
\caption{{Entanglement entropy as a function of timestep in model C, for  two different values of the smallness parameter $\epsilon$.}}
\label{fn6m}
\end{center}
\end{figure}

The numerical results for the entanglement entropy are presented in fig.\ref{fn6m}, for $\epsilon=0.2$ and for $\epsilon=0.4$. Note that $\epsilon=0.4$ is quite a large perturbation, given that there are three allowed deformations of the pair state away from the leading pair state $S^{(1)}$, all proportional to $\epsilon$. But in each case we again find that the entanglement monotonically increases with $n$.

\section{Entanglement entropy for `burning paper'}\label{burn}

It is important to note the difference between the behavior of entanglement entropy found for  the back hole and that found when we burn a piece of paper. In the latter case the photons leave from the surface of the paper, and are thus typically in an entangled state with the 
part of the paper they leave behind. As Page argued \cite{page}, the entanglement entropy between the radiation and the paper at first goes up, but after about half the paper has burnt away, the entanglement entropy starts going down again, reaching zero when the paper is completely converted to its radiated products. The reduction in entanglement happens for the following reason. A radiated photon may be in an entangled spin state with the atom which radiated it, but eventually this atom will also be radiated away as ash, in order for the paper to completely disappear. Then the entanglement will be between two parts of the radiated energy: the radiated photon and the radiated atom. (In general interactions among the many atoms and photons in the paper will make the overall entangled state quite complicated.) The information in the original paper is thus encoded in correlations among the radiation products, which are in an overall pure state; they are not entangled with anything other than themselves.

In this section we make a numerical model of `burning paper',\footnote{We are grateful to Don Page for several discussions on toy models of this type.} and observe that the entanglement entropy in this case indeed goes up and then returns to zero. The essential difference between the black hole and the case of burning paper was explained in \cite{mathurfuzz}. We discuss this difference again, and also provide a laboratory model which is more analogous to the black hole evaporation situation.

\subsection{A model of burning paper}\label{subfour}

We do not distinguish between photons and atoms. We take a set of $N$ objects, which we call atoms for simplicity. We allow each atom to have two spin states, $\u$ and $\d$. The evolution is described as follows:

\b

(i) The initial state (timestep $n=1$) is a wavefunction for the $N$ spins. Let us take it to be a definite but arbitrarily chosen set of spins, for example if we have $N=11$ then we could take the state at timestep $n=1$ to be
\be
|\Psi\rangle_0=\d\u\d\u\u\d\d\u\d\u\d
\label{ex4}
\ee

\b

(ii) We imagine that the radiation takes place from the right side of this set of spins (thus the left side is to the `interior' of the paper). To get the state at timestep $n=2$ we look at the two atoms on the rightmost end. Only one of these atoms will be radiated, but the actual radiation process depends on the state of both these two spins. The rules for radiation are chosen as follows:

(a) If both atoms are $\u$, then the radiated atom is $\u$ and the other $\u$ atom stays behind as the rightmost atom in the set.

(b) If both atoms are $\d$, then the radiated atom is $\d$ and the other $\d$ atom stays behind as the rightmost atom in the set.

(c) We also add rules for the case that the two rightmost atoms are up-down or down-up:
\be
\u\d~\r~\sqi \Big ( \u \Big | \d + \d \Big | \u\Big )
\ee
\be
\d\u~\r~\sqi \Big ( \u \Big | \d - \d \Big | \u\Big )
\ee

In this way we generate an entanglement between the radiated atom and the surface of the radiating paper. We draw a vertical bar to separate the atoms which have been radiated and the atoms which are left behind. In the example (\ref{ex4}) the two rightmost spin were $\u\d$, so we will get 
\be
|\Psi'\rangle_2=\sq \d\u\d\u\u\d\d\u\d\u\Big |\d+\sq \d\u\d\u\u\d\d\u\d\d\Big |\u
\label{ex4p}
\ee
This state is not yet the final state at timestep $n=2$, so we have denoted it with a prime.

\b

(iii) In a piece of burning paper we also have a Hamiltonian which makes the atoms in the paper interact with each other.  Thus the entanglement between the radiation and the {\it surface} of the paper changes  to an entanglement between the radiation and the {\it bulk} of the paper. In our model we take the following rule to incorporate such an effect. After every step of radiation, we perform a clockwise rotation on the spins left behind in the paper; i.e., the atom at the rightmost end is moved to the leftmost position, and the other atoms slide one step each to the right. Thus for the example that we have taken, we get
\be
|\Psi\rangle_2=\sq \u\d\u\d\u\u\d\d\u\d\Big |\d+\sq \d\d\u\d\u\u\d\d\u\d\Big |\u
\label{ex4pp}
\ee
This will be our state at timestep $n=2$. The evolution process now repeats: we look  at the last two spins just before the vertical bar, use the rules (a)-(c) to eject another spin, perform the clockwise permutation, and then move to the next step.

\b

(iv) At timestep $n+1$, we will have $n$ atoms to the right of the vertical bar and $N-n$ atoms to the left of the bar. The overall state is entangled between these two sets of atoms. We trace over the atoms in the paper (i.e. the atoms to the left of the vertical bar) to get a reduced density matrix $\rho$ describing the atoms that have been radiated. Finally, we compute
\be
S_{ent}(n)=-Tr \rho\ln \rho
\ee
It is clear that $S_{ent}$ vanishes at $n=1$ (since there are no radiated atoms). It is also clear that $S_{ent}$ will have to vanish at the final timestep $N+1$ since all atoms are to the right of the bar, and there is nothing left for them to entangle with. At timestep $N$ there is only one spin left in the paper, and we must have $S_{ent}\le \ln 2$. Thus we expect the kind of behavior noted by Page \cite{page} for all normal radiating bodies: an increase in $S_{ent}$, a turning point when about half the degrees of freedom have been radiated, and a return to $S_{ent}=0$ when the burning process is completed. 

\begin{figure}[htbp]
\begin{center}
\includegraphics[scale=.58]{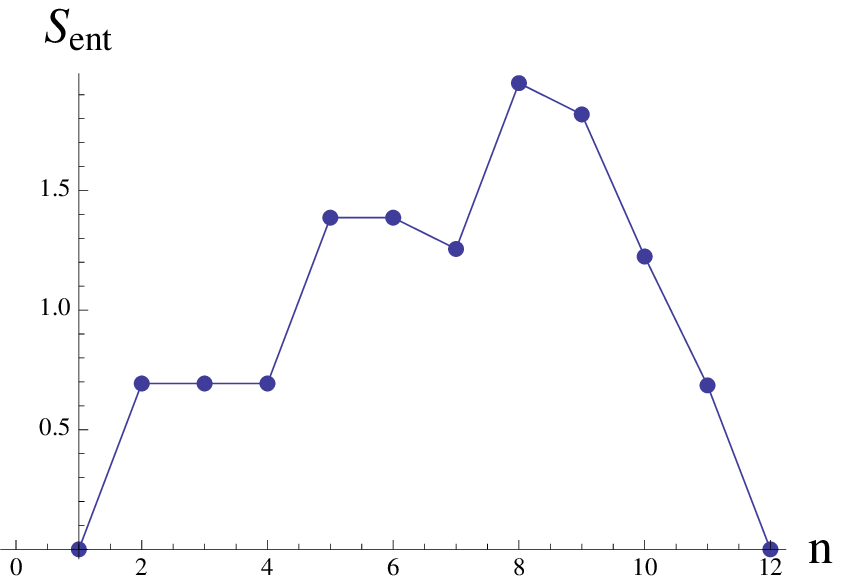}\hskip 1 true in
\includegraphics[scale=.58]{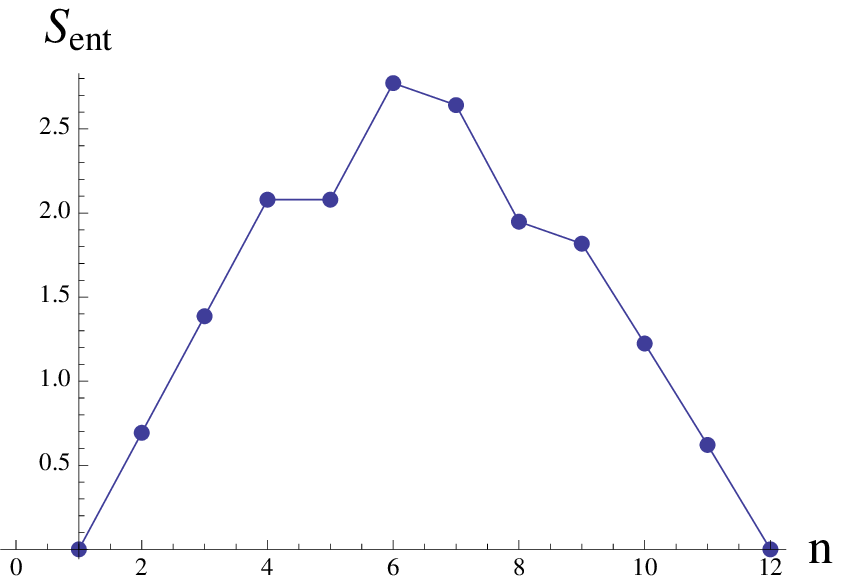}
\caption{{Entanglement entropy for a model of `burning paper'. The first graph corresponds to the initial state $\{\d\u\u\d\u\d\d\u\u\u\d\}$ and the second to $\{\d\u\d\d\d\u\d\u\d\d\u \} $.}}
\label{fn7m}
\end{center}
\end{figure}

Numerical results from this model are given in fig.\ref{fn7m}. For two different choices of initial state, we plot $S_{ent}$, as a function of the timestep. The first case starts with the spins $\{\d\u\u\d\u\d\d\u\u\u\d\}$ while the second starts with 
$\{\d\u\d\d\d\u\d\u\d\d\u \} $. As expected \cite{page}, we find that $S_{ent}$ at first rises, and after roughly the halfway point, decreases, ending at zero when the burning process is completed.

\subsection{A physical model resembling black hole evaporation}

We can now see how the models of black hole evaporation studied in section \ref{secthree} differ in an essential way from models of burning paper. For the model of burning paper, the state generated at any timestep $n$ depends on the state of the atoms near the surface of the paper; such a dependence is encoded in the rules (a)-(c) of section \ref{subfour}. The state of the radiated atom and the atom left behind at the surface is typically an entangled one, and chosen  from  a vector space spanned by a few possibilities. For information to be carried out in the radiation, it is important that this vector space has dimension greater than one. To see this fact, consider a case where this vector space was in fact 1-dimensional. For instance, we can let the `paper' be made of $N$ identical spinless bosons, all in the same state.  We let one boson be emitted at each timestep, in the same state each time. Now there will be no entanglement between the emitted bosons and the bosons left behind. But in such a situation there will be no information in the initial state to carry out, and there will be no information in the final state of radiated bosons either. By contrast in the model of burning paper that we have taken in section \ref{subfour}, the evolution at any timestep at the surface of the paper took one of three different forms, with the probabilities of the three forms being order unity each; i.e. it was {\it not} the case that two of the possibilities had infinitesimal amplitudes ($O(\epsilon)$, with $\epsilon\ll 1$) while the third had amplitude satisfying $|{\cal A}|^2=1-O(\epsilon^2)$. The paper could be in one of $2^N$ states, and because the radiation process at each step depended in a nontrivial way on the state of the paper near its surface, the radiation could faithfully carry out information about the choice of initial state, in a scrambled form.

Let us now look again at the case of black hole evaporation, assuming as we have done in section \ref{secthree} that we have a traditional horizon and therefore the state at the horizon is close to the vacuum in the `good slicing'. Now we create particle {\it pairs} at this horizon, and these pairs are in an entangled state. But the state of the pair is chosen from an effectively 1-dimensional vector space; i.e. it is always the state $S^{(1)}$ (eq.(\ref{pairs})), upto small corrections from other states. Note that unlike the model of spinless bosons above, this time the radiated quanta entangle with the quanta left behind. It is the state of the {\it pair} which is taken from an effectively 1-dimensional vector space. This effectively  1-dimensional nature of the vector space  creates the entanglement problem: at leading order we get the same increase in entanglement each time, since the evolution is not sensitive to any detail of the hole. The small corrections only reduce the entanglement from its leading order value by a small amount, as proved in \cite{mathurfuzz}. 

\subsubsection{A laboratory model}

\begin{figure}[htbp]
\begin{center}
\includegraphics[scale=.25]{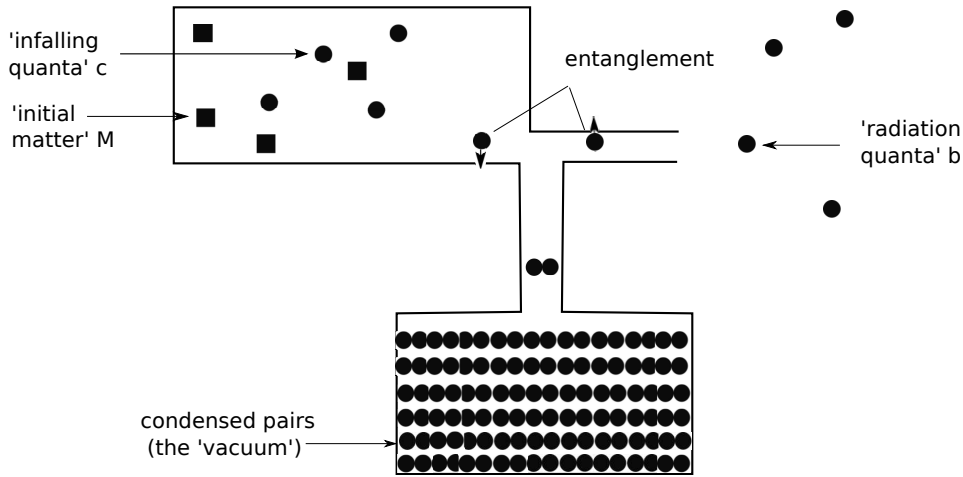}
\caption{{A laboratory model capturing some features of black hole evaporation. Condensed electron pairs reside in the tank at the bottom; one pair at a time rises up in the tube and disassociates into an entangled pair of electrons. One member of the pair escapes as radiation, while the other is collected in the box. (This box also contains a set of quanta to represent the initial matter which made the hole). Until the pairs are exhausted (or the collection box becomes too full), the entanglement entropy between the radiation and the box keeps rising.}}
\label{fn3}
\end{center}
\end{figure}

Let us make a laboratory model that will mimic the features of such black hole evaporation. We depict this model in fig.\ref{fn3}. We have a tank at temperature close to zero, filled with `electron pairs'; these are entangled pairs of electrons in the state
\be
\sq(\u\d-\d\u)
\label{state5}
\ee
Since each pair is bosonic, we imagine the quanta in the tank to be a bose condensate of these pairs. Thus there is a unique state for the entire collection of quanta in the tank. In our analogy with the black hole
the store of pairs in the cold tank will represent the vacuum, from which particle pairs can be pulled out.

At each timestep of evolution, one electron pair  evaporates up from the tank into a chamber, where it disassociates into its component electrons. One electron flies out to the right as `radiation', while the other is collected in a box on the left. Note that this produces an entangled state of the form (\ref{state5}) between the radiated quantum and the quantum collected in the box. 

To complete the analogy with the black hole, we imagine that the box on the left starts out with a collection of $M$ atoms; these represent the initial matter which made the black hole. Consider the creation of any new pair from the condensate in the tank. There will be some slight change in the state of this pair, due to the  $M$ atoms, and due to electrons already collected in the box from previous steps. Such small effects can arise for example from lattice vibrations caused by collisions of these particles with the walls of the box; such collisions will indirectly induce small vibrations of the chamber where the pair was disassociating. But we assume that the net change of the state of the newly created pair from all such effects is small, governed by a small parameter $\epsilon\ll 1$. We will have this situation if the box is large, and we study the evolution only upto a time when it is not so full of quanta that these quanta spill over and occupy the disassociation chamber; this requirement is analogous to asking that we study the black hole evaporation process only for times where the size of the horizon remains in the semiclassical domain.

\subsubsection{Comparing the laboratory model with the black hole}

To summarize, with the model set up as above, we have captured most of the features of black hole evaporation in the case where we assume a traditional horizon. Particle pairs are pulled out of our reservoir tank; each new pair is pulled out in the same state and gives rise to an entangled pair upon disassociation into a radiated electron and an electron collected in the box. There are small corrections to this entangled state from atoms initially placed in the box, and also from electrons collecting in the box, but these effects are assumed small. In this situation the entanglement between the radiated quanta and the system left behind will keep rising. This monotonic rise is different from what we got for the case of burning paper, where the entropy behind to decrease after a point. Of course in this laboratory model we cannot keep on creating pairs indefinitely; when the tank is empty the pair creation will stop. In the traditional picture of the black hole, on the other hand, the pair creation will continue as long as the horizon has a semiclassical size. The goal of our laboratory model of condensed pairs
is to just to illustrate the nature of the pair creation process, not to map the black hole to a laboratory situation. 

Of course there is one feature of black hole evaporation that we cannot capture in such a model: the situation at the endpoint of evaporation. In our model of disassociating electron pairs we can stop the evolution after some timestep $N$, perhaps when all the pairs in the tank are exhausted. At this stage there is a large entanglement ($S_{ent}\approx N\ln 2$) between the radiated quanta and the box, but this is not a contradiction in any way; if there are many electrons in the box then we can certainly have a radiation state that is highly entangled with this box. 
The novel thing about black holes is the attractive nature of gravity, which can reduce the net energy of a system to a value arbitrarily below the rest energy of its constituents. In the box above we have the atoms representing  the initial matter $M$ in the hole, and we have the electrons that represent the quanta $\{ c\}$ that fall into the hole. In the laboratory model the energies of these two kinds of matter add up. But in the black hole the $\{ c\}$ quanta have negative energy because of the gravitational attraction of the hole, and after enough $\{ c\}$ quanta have collected, the net mass of $M$ and $\{ c\}$ reduces down to planck scale. It is this latter fact which creates the information problem: either such small mass states (remnants) must possess an infinite degeneracy (to allow for arbitrarily large entanglement) or the hole disappears completely, leaving the radiation in a mixed state with nothing that it is `mixed with'. 

\subsection{Summary}

The above models of `small corrections to Hawking radiation' (section \ref{secthree}) and the model of burning paper above illustrate the essential power of the inequality proved in \cite{mathurfuzz}: the Hawking evolution is stable against small corrections, in the sense that such small corrections cannot remove the large entanglement between the radiation and the hole found at leading order. We see explicitly that the case of `burning paper' is essentially different; for this latter situation we indeed reproduce the conventional expectation discussed by Page \cite{page}. 

In our numerical models we were able to pursue the evolution only for about $n=12$ timesteps. The number of states of the system rises exponentially, and limitations of computer memory and time do not allow us to go much further. It is likely that the large space of states can be handled by techniques like Monte Carlo, as in the work on lattice gauge theory, but we have not tried to do this since the issue to be understood here was the role of `delicate correlations', and a computation using approximations would not be convincing in its conclusions. In principle there is no barrier to evolving through more timesteps if more computational power is available.

\section{Infall into fuzzballs}

\begin{figure}[htbp]
\begin{center}
\includegraphics[scale=.25]{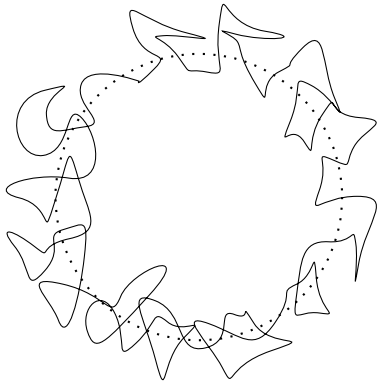}
\caption{Schematic diagram of a `fuzzball'. The geometry far from the horizon region is the same as that of the black hole, but as we approach the place where the horizon would have been in the traditional picture,  we find a complicated quantum gravitational solution, different for every microtstate of the hole.}
\label{ftwop}
\end{center}
\end{figure}

The examples above illustrate the general theorem of \cite{mathurfuzz} that small corrections to Hawking's computation cannot solve the information paradox. We now turn to how the paradox is actually resolved in string theory. One finds that the corrections to the evolution of low energy modes at the horizon are {\it not small}; in fact the entire structure of the hole at the horizon is radically different from the traditional black hole geometry \cite{fuzzball1,fuzzball2}. The details of the structure depend on the choice of microstate (from among the $Exp[S_{bek}]$ states of the hole). Such microstates are called `fuzzballs', because the generic state of the hole is a very nontrivial quantum fuzz in the region where the horizon would have been, instead of the smooth vacuum of the traditional black hole geometry (fig.\ref{ftwop}).

Changing the state of the hole at the horizon certainly resolves the information paradox; if we no longer have the repeated `pair creation from vacuum' that we discussed in section  \ref{ineq}, then we evade the whole problem of monotonically growing entanglement between the radiation and the hole. In fact for simple fuzzballs the radiation has been computed explicitly, and seen to lead to no information problem. The fuzzball has no horizon, but it does have an ergoregion, and radiation emerges by the process of ergoregion emission \cite{ross,myers}. This process also creates quanta in pairs, but the inner member of the pair does not fall into a horizon; instead it resides in the ergoregion, where it affects the evolution of later quanta, and in fact can emerge itself (in general after severe interactions) at a later time. This is exactly the behavior of radiation from `burning paper', discussed in section  \ref{burn}. One finds that the {\it rate} of radiation agrees exactly with what would be expected for Hawking radiation from this microstate\cite{cm1} ; this agreement is based on computing the radiation process in the CFT where the radiation can be computed for any microstate. But while the rate is the same as that for the Hawking process, the mechanism of emission is different, and does not lead to the information problem.

Thus we resolve the information paradox, but we can now wonder about the `infall problem'. In \cite{mathurrecent} the difference between these two issues was discussed in detail, and it was noted that they are in a sense {\it opposite} problems. The information paradox asks about the detailed state of the low energy quanta ($E\sim kT$) emitted by the hole, and how {\it different} states of the hole can  give rise to {\it different} states of the radiation. The `infall problem' asks if there is an effective description in which these various microstates behave in the {\it same} way for the purpose of describing  the infall of heavy ($E\gg kT$) objets over short times (times of order the crossing scale rather than the Hawking evaporation scale).  In particular we can ask if this effective coarse grained dynamics somehow involves the traditional black hole geometry which has a smooth horizon. Some initial comments about the infall problem in the fuzzball picture were given in \cite{mathurrecent}; here we consider this issue in more detail. 

\subsection{Rindler spacetime}

Let us begin by asking what fuzzballs suggest about the behavior of Rindler space, which shares many features with black holes. 

\begin{figure}[htbp]
\begin{center}
\includegraphics[scale=.85]{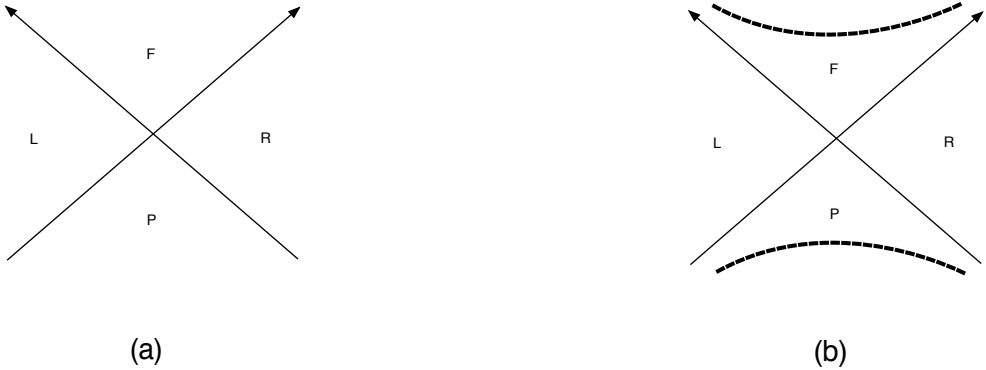}
\caption{{(a) Rindler space (b) The Penrose diagram of the extended Schwarzschild hole. The region near the intersection of horizons is similar in the two cases.}}
\label{fn5}
\end{center}
\end{figure}

Fig.\ref{fn5}(a) depicts Minkowski space, with its four Rindler quadrants:
R (right), L (left), F (future), P (past). The Minkowski space metric is
\be
ds^2=-dT^2+dX^2
\ee
Under a coordinate change
\be
T=r\sinh t, ~~X=r\cosh t
\ee
we get the metric in the right quadrant
\be
ds^2=-r^2 dt^2+dr^2
\label{ssone}
\ee
In fig.\ref{fn5}(b) we depict the extended Schwarzschild geometry of the black hole. The region near the center of the diagram (where the two horizons intersect) looks just like the diagram of Minkowski space broken into its Rindler quadrants. The R (right) region is now the exterior of the black hole, and the analog of (\ref{ssone}) is the metric
\be
ds^2=-(1-{2M\over \t r})d\t t^2+{d\t r^2\over 1-{2M\over \t r}}+\t r^2d\Omega_2^2
\label{sstwo}
\ee
For ${\t r-2M\over 2M}\ll 1$, we write
\be
 t={\t t\over 4M}, ~~~ r=2\sqrt{2M}\sqrt{\t r-2M}
\ee
and we get from (\ref{sstwo})
\be
ds^2\approx - r^2 d t^2  + d r^2+ (2M)^2 d\Omega_2^2
\ee
giving Rindler space (\ref{ssone})  times a sphere.

Note that if we go to the Euclidean section in the Rindler metric (\ref{ssone})
\be
ds^2=r^2d\tau^2+dr^2
\ee
then regularity at $r=0$ would need the compactification
\be
\tau\sim \tau+2\pi
\ee
so that we get the temperature ${1\over 2\pi}$ in the Rindler coordinate frame $(\tau,r)$ \cite{unruh}. At any radius $r_0$, we can go to the local orthonormal frame where we measure time as
\be
ds^2\approx d(r_0\tau)^2\equiv d\tau'^2
\ee
The temperature measured in time $\tau'$ is ${1\over 2\pi r_0}$, so we see that this temperature diverges as we go towards the horizon $r\r 0$.

This is all a standard discussion of Rindler space, but we now make a series of observations that relate the observations above to fuzzball solutions:

\b

\b

(i) Consider the Rindler metric (\ref{ssone}). Suppose the spacetime carries  a scalar field $\phi$. Then we will see a thermal bath of quanta of $\phi$ with temperature ${1\over 2\pi r_0}$ at radius $r_0$ \cite{unruh}.

\b

(ii) Suppose this scalar field is an interacting one, with Lagrangian $\h\p\phi\p\phi-\lambda\phi^4$. Then the $\phi$ quanta in this thermal bath will be interacting among themselves with the quartic interaction $\lambda \phi^4$ \cite{unruh2}.

\b

(iii) One field that is always present in any theory of spacetime is the gravitational field, whose fluctuations we will term $h_{\mu\nu}$. Thus we always expect a thermal bath of gravitons in the Rindler frame.

\b

(iv) Following (ii) above, this thermal bath should consist of {\it interacting} gravitons, so in the Rindler frame we should see fully non-linear gravity near the horizon. This suggests that the Rindler problem needs to include the physics of curved spacetime to be consistently defined. Since we are looking at a quantum problem, we really mean here solutions of the quantum gravity theory (which would be string theory in what follows).

\b

\begin{figure}[htbp]
\begin{center}
\includegraphics[scale=.75]{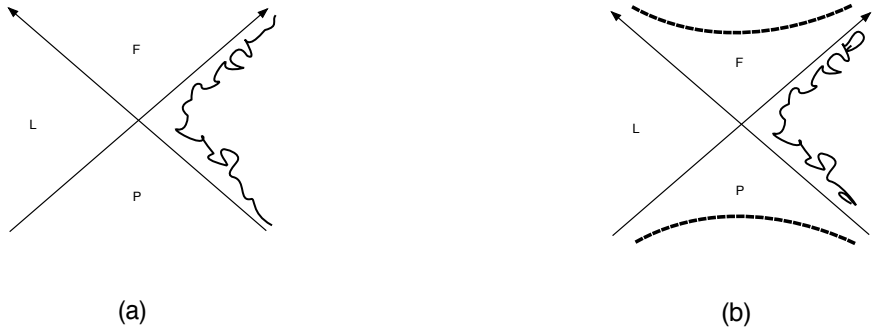}
\caption{{Fuzzball structure for Rindler space (a) and the black hole (b). While the extended Schwarzscild metric has four regions separated by horizons, the fuzzball solutions of string theory occupy only the right (R) quadrant, and end in a quantum fuzz near the place where the horizon would have been (figure (b)). Carrying this notion over to Rindler space, we should think of the individual states of the R quadrant as ending in a quantum gravitational fuzz near the horizon (figure (a)).}}
\label{fn11}
\end{center}
\end{figure}

(v) We can now wonder what are the nonlinear gravity solutions that should appear near the horizon. In fig.\ref{fn11}(b) we depict the full Penrose diagram of the Schwarzschild metric (\ref{sstwo}), and also what we find in the fuzzball picture: the Schwarzschild like metric of the right (R) region is severely modified near the horizon and ends in a `quantum fuzz'.  By analogy, it appears natural to conjecture that a similar situation would hold for Rindler space, and we depict this in fig.\ref{fn11}(a): the geometry of the right Rindler region (R) ends in a quantum fuzz; this quantum state is supposed to describe the interacting gravitational field of (iv) above. The black hole has a finite area horizon and thus a finite number of degrees of freedom, while the Rindler solutions should to be thought of as fuzzball solutions with some kind of $M\r\infty$ limit in mind. 

\b

Thus we are conjecturing that the Rindler observer sees gravity solutions that {\it end without horizon} at the boundary of the 
right Rindler region. This may seem odd, since in Minkowski coordinates the spacetime continues past a smooth horizon into the other Rindler quadrants. But note that the Minkowski spacetime was flat, while the steps (i)-(v) in the above argument have force us to think of Rindler space as a very nontrivial quantum gravitational solution near the horizon. We will now discuss how these two different perspectives can be reconciled.

\subsection{The eternal hole}

There have been many observations relating the extended Penrose diagram of the black hole to the idea of entangled systems. We recall these ideas briefly, and then discuss how one might understand the role of fuzzball solutions in the context of these ideas. This discussion will lead us to a version of `complementarity' for the black hole.

\b

(a) Israel \cite{israel2} noted that the left and right sides of the eternal black hole Penrose diagram (fig.\ref{fn5}(b)) can be regarded as the two copies of a system that arise in the thermo-field-dynamics (TFD) formalism \cite{umezawa}. Thus the overall state of the system would be an entangled state of the fields on the left and right sides, and either side alone would have to be described by a density matrix where the other side was traced over.

\b

(b) Maldacena \cite{maldacena2} observed that the dual to the eternal black hole (in $AdS$) should be a pair of CFT's, with one copy of the CFT corresponding to each asymptotic infinity of the full Penrose diagram.\footnote{See also \cite{horowitzmarolf,trivedi}.} The states of these two CFTs are entangled, in accordance with the entanglement of the  bulk in (a) above.\footnote{The notion that the ensemble sum over states should be replaced by the two sides of a black hole diagram has also been used recently in \cite{sen}.} 

\b

\begin{figure}[htbp]
\begin{center}
\includegraphics[scale=.85]{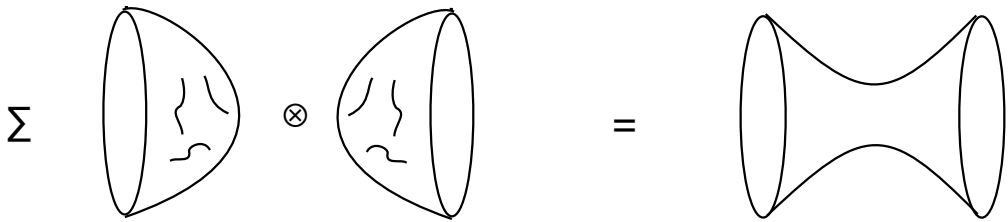}
\caption{{An entangled state of gravity solutions can be represented as a single {\it connected} gravitational solution.}}
\label{fn6}
\end{center}
\end{figure}

(c) Van Raamsdonk used the above notions to  taken the issue of entanglement further \cite{raamsdonk}. He observes that states in the individual CFTs would correspond to a gravity solution with only one asymptotic infinity, and it is the entanglement over two sets of such gravity solutions that must generate the `bridge' that links the two sides of the Penrose diagram of fig.\ref{fn5}(b). We depict this general notion in fig.\ref{fn6}. Summing over the pair of gravity solutions on the LHS (with appropriate weights) gives an entangled state of gravity which can be alternatively represented by a manifold which has a `bridge' linking the two sides.

(d) At this point we need to consider the  nature of the states involved in each factor of the sum in (c). In the traditional picture of the black hole we would have two kinds of states: states like $AdS$  plus a few  quanta (which have no horizon), and states that have a black hole (these do have a horizon). The latter states would be more numerous (because of the high entropy of the hole).  For the black hole states we would need to draw something different from the geometries  on the LHS of  fig.\ref{fn6} to take into account that each of the left and right factors on the LHS itself possesses a horizon.  

\begin{figure}[htbp]
\begin{center}
\includegraphics[scale=.85]{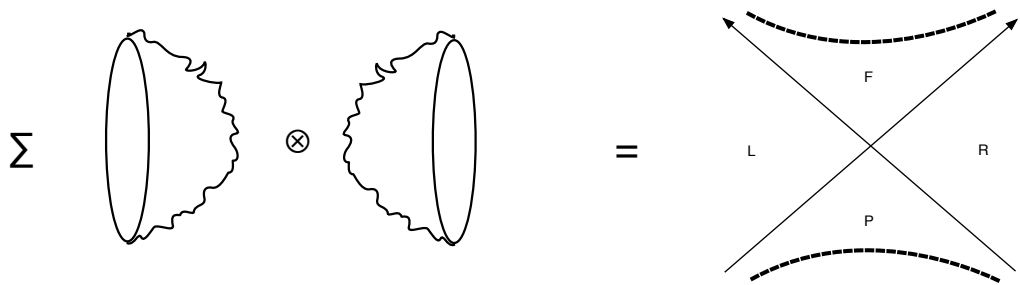}
\caption{{Black hole microstates are fuzzballs that `end' without forming a horizon. Summing over pairs of microstates (with appropriate weights) should give the geometry of the extended Schwarzschild hole.}}
\label{fn7}
\end{center}
\end{figure}

But in the fuzzball picture we do not have two different classes of states; all states are horizon-free and simply differ from each other in their degree of `complexity'. Thus we should write the sum as in fig.\ref{fn7}: the energy weighted sum over all microstates gives the eternal hole geometry which connects the two disconnected factors.\footnote{\cite{raamsdonk}  postulates the relation depicted in fig.\ref{fn7}. Here we are simply noting that if the states were split into `thermal $AdS$ states' and `black hole states' (with horizons) then it is not clear what such a relation would mean.}

\b

\begin{figure}[htbp]
\begin{center}
\includegraphics[scale=.85]{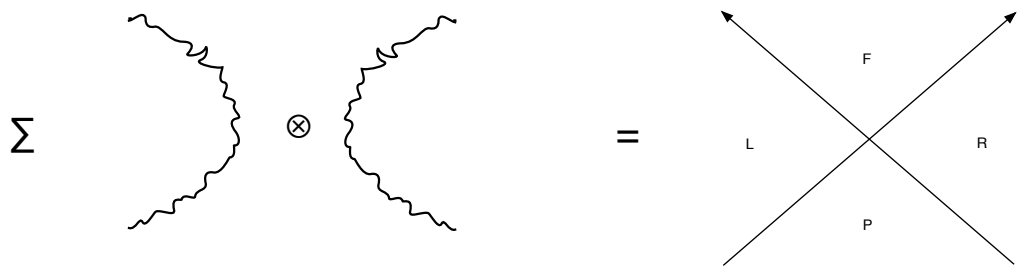}
\caption{{Extending the relation in fig.\ref{fn7} to Rindler space. A (weighted) sum over products of gravitational solutions (residing in only one quadrant each and having a very quantum gravitational region  near the horizon) should give smooth Minkowski spacetime.}}
\label{fn12}
\end{center}
\end{figure}

Taking the limit of $M\r\infty$ we should recover an analogous statement for Rindler space: a weighted sum over Rindler wedges `ending without horizon' should give Minkowski spacetime (fig.\ref{fn12}). 

\b

(e) With this extended geometry we can pass smoothly through the horizon. In the simpler Rindler case, the region F is described by field modes that are left moving in the R sector and right moving in the L sector. (These modes simply extend across their respective horizons to give a compete basis in the F region.) Similarly, in the eternal black hole geometry the modes inside the future horizon are given by continuing appropriate modes from the R and L regions. 

\b

(f) Thus a single fuzzball does not have a horizon that we can continue through to reach the `interior' of the black hole. But the ensemble sum over fuzzballs can be given such an extension. Now consider the conjecture on the infall problem made in \cite{mathurrecent}. It was noted that infalling objects whose physics should not be affected materially by Hawking radiation have energies $E\gg kT$. For these high energies we can replace correlation functions in a given fuzzball by the corresponding  correlation function in the ensemble over fuzzballs (assuming that the fuzzball is a generic one). Since this ensemble sum is described by a geometry which can be continued across the horizon, we obtain an effective description for infalling observers where nothing happens at the horizon.

Let us make this proposal more explicit. In any statistical ensemble, there exist `generic' observables $\hat O_i$, whose correlation functions in a {\it generic} microstate are equal, to a good approximation, to the {\it ensemble average} over microstates:
\be
\langle \Psi|\hat O_1\hat O_2|\Psi\rangle\approx \sum_i e^{-\beta E_i} \langle \Psi_i|\hat O_1\hat O_2|\Psi_i\rangle
\label{ppone}
\ee
The RHS can be written as
\be
\sum_i e^{-\beta E_i} \langle \Psi_i|\hat O_1\hat O_2|\Psi_i\rangle= Tr~\Big (\rho_\beta \hat O_1\hat O_2\Big )
\ee
where $\rho_\beta=\sum_i e^{-\beta E_i }|\Psi_i\rangle \langle \Psi_i |$. 
Let $|\Psi\rangle$ in (\ref{ppone}) be a generic fuzzball state; this state has no horizon. The correlator can be approximated by one in the thermal density matrix, which, as we have argued following \cite{israel2,maldacena2,raamsdonk}, is given by the eternal black hole spacetime. It is only in this sense that we recover a geometry with a horizon: the indiviual fuzzball has no horizon, while the eternal black hole spacetime does. But note that after averaging we lose all information about the initial microstate $|\Psi\rangle$. Thus we cannot hope to recover the details of $|\Psi\rangle$ from the Hawking radiation observed in the ensemble average. And indeed, the eternal black hole diagram does not have the spacetime structure to radiate any information. In fact there is no unique choice of matter vacuum  that goes along with its geometry; we can choose the vacuum so that it radiates energy, absorbs energy, or does neither. 

The only question  that remains is: what operators $\hat O_i$ are `generic' in the sense that they satisfy (\ref{ppone}). It is natural to conjecture that these will be operators that are natural for measurements in the frame of a lab falling from infinity to the surface of the fuzzball: the high relative velocity between the lab and the fuzzball makes these operators `high impact' in the sense that they will cause the fuzzball to jump from one state to a linear combination of states in a {\it band} around the initial energy of the fuzzball. This band has a large number of states because of the high entropy of the black hole, and the smearing over this band makes the correlators insensitive to the precise choice on initial microstate $|\Psi\rangle$ in (\ref{ppone}).

\b

(g)  Modes of the Hawking radiation itself are low energy ($E\sim kT$), and for these we cannot make the approximation (\ref{ppone}) of using a thermal ensemble. (In \cite{mathurrecent} the following analogy was used. The evolution of Hawking quanta is like the Brownian motion of a single atom in a gas, sensitively dependent on the detailed positions of the gas molocules, while $E\gg kT$ observers are like heavy objects moving through the gas, described by an overall viscosity that is computable in an ensemble average.) Thus the fuzzball structure of microstates provides the order unity correction to the evolution of Hawking quanta (which is necessary to resolve the information paradox), and the ensemble average description appropriate to infalling observes gets modeled by an effective geometry that continues smoothly through a horizon. 

\subsection{Comment on the relation depicted in fig.\ref{fn6}}

\begin{figure}[htbp]
\begin{center}
\includegraphics[scale=.75]{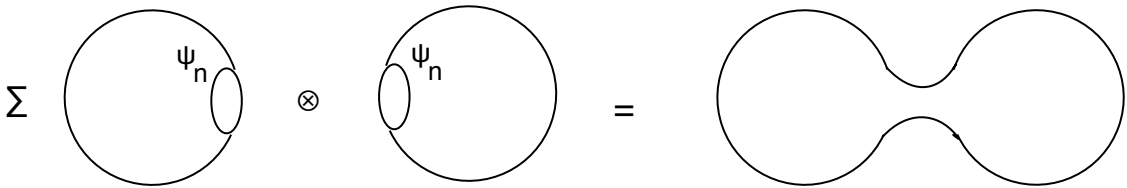}
\caption{{The `sewing rule' in 2-d Eucliean CFT. If we sum over states entangling two spheres  then we generate get a handle between the spheres, giving the CFT on a single connected manifold.}}
\label{fn8}
\end{center}
\end{figure}

Let us see why a relation of the kind in fig.\ref{fn6} might be reasonable. Consider a Euclidean CFT on a 2-d manifold. Start with this CFT on two spheres as on the LHS of  fig.\ref{fn8}. On each sphere, cut out a hole given by $|z|<1$, in some coordinate chart $z$. Let $|\psi_n\rangle$ be the state at the edge of each hole. Consider the tensor product theory of the two spheres, and sum over all states $|\psi_n\rangle$ with  a weightage factor $e^{-(\tau h+\bar \tau \bar h)}$ (here $h, \bar h$ are holomorphic and antiholomorphic weights of the state, and $\tau$ is a modular parameter). This process generates a handle connecting the two spheres, so that at least in this case we may replace the complicated sum over states by a smooth manifold. 

\begin{figure}[htbp]
\begin{center}
\includegraphics[scale=.75]{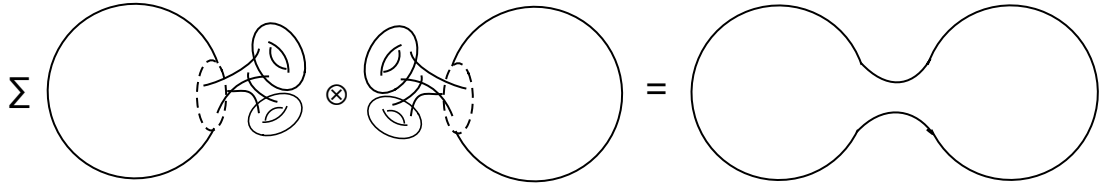}
\caption{{The states $|\psi_n\rangle$ in fig.\ref{fn8} give a complicated structure when continued inside $|z|=1$.  Thus an entangled sum over pairs of complicated geometric states generates the smooth manifold on the RHS.}}
\label{fn9}
\end{center}
\end{figure}

Note that on each individual sphere, in any one state $|\psi_n\rangle$, we would not see the smooth and  circularly symmetric handle if we try to pass  through the circle $|z|=1$ to smaller values of $|z|$. Suppose we took the CFT to be that of 2-d gravity (described, say,  by dynamical triangulations). Then continuing the state inside $|z|=1$ would give in general a complicated geometry for each $|\psi_n\rangle$. This complication can be manifested in suitable correlation functions by making one or more points in the correlator approach the circle $|z|=1$. We now have the situation in fig.\ref{fn9}, where a weighted sum over complicated geometries reproduces the smooth geometry which has a handle connecting the two sides. The individual states on any sphere are akin to fuzzball states, and the weighted sum is akin to the eternal black hole diagram. In short, the relation of fig.\ref{fn9} suggests that we should think of `fuzzball microstates' as a  `complete set' of gravitational states that we can use in sums of the form $\sum_n e^{-\beta E_n}\sum_n |\psi_n\rangle\langle \psi_n|$ for generating `handles' on spacetime.\footnote{The singularities in the black hole Penrose diagram appear to be related to the limited thickness of the `handle' when $M$ is finite; it would be good to understand this better.}

\subsection{Complementarity}

't Hooft postulated the notion of black hole complementarity \cite{thooft1}, and Susskind and others developed this notion further \cite{suss1}. The essential idea is that the black hole physics has two different descriptions. For an observer outside the horizon, we should use a description where there is no region interior to the horizon: objects fall to the horizon, with this infall slowing down because of the diverging redshift. For this outside observer, the infalling object gets destroyed at the horizon and its information should be encoded in the emerging Hawking radiation. For an infalling observer, on the other hand, nothing special happens at the horizon, and he carries his information with him into the region behind the horizon. The existence of two descriptions is supposed to be consistent because of the kinematical structure of the black hole spacetime: the observer falling into the hole cannot send signals out, and lives for only a short time before hitting the singularity. These two descriptions of the hole are `complementary' in the sense that any attempt to describe the physics seen by the outside observer and the infalling observer at the same time would amount to `double counting'. 

While complementarity is an appealing idea, and may bypass the information problem, it runs into an immediate difficulty. Consider the full Penrose diagram of the hole foliated by the nice slices depicted in fig.\ref{fthree} (and in more detail in fig.\ref{ftwo}). This foliation captures 
the infalling shell which made the hole, as well as all the $\{c, b\}$ quanta that are emitted in the Hawking radiation process. Since the geometry is regular everywhere, there is no reason why physics on one part of these slices should give a complementary description of the physics on other part of the slice; the entire slice would appear to be the correct Cauchy surface. There seems to be no way to build in complementarity in the black hole geometry while keeping physics unchanged in all other examples of evolution in gently curved spacetime.

But with the recognition that black hole microstates are fuzzballs (and therefore not approximated by the Penrose diagram of fig.\ref{fthree}) we can take another look at the issue. For the outside observer, we consider just one fuzzball state (depicted in fig.\ref{ftwop}), and indeed in this case physics `ends' in the vicinity of where the traditional horizon would have been.  For the infalling observer, we have to ask what we mean by his `observations'. Are these expectation values in one fuzzball state or in the ensemble average? If we take the latter, then as discussed above we can continue his dynamics past the horizon into an extended spacetime, and in particular find  the `interior' of the hole. Thus in this approach the interior of the hole arises as an ensemble averaged description of the collective modes of fuzzballs. It would be interesting to relate this to other approaches where the microstates are averaged to obtain an effective geometry \cite{bala}. 

It is also interesting to recall the idea of how one may see a `brick wall' just outside the horizon  \cite{thooft1}. The idea may be summarized as follows.  The standard Hawking radiation process produces radiation in a complicated entangled state of $\{ c, b\}$ pairs. But suppose we choose a particular state of the radiated $\{ b\} $ quanta. This particular state if quite different from the full entangled state, and if we follow the $\{ b\}$ quanta back towards the horizon in this state then they will not meet up with the appropriate $\{ c\}$ quanta to create the vacuum state that Hawking used. In fact due to the large blueshift in this backwards evolution, this non-vacuum state will have a very large stress-energy near the horizon, and we can imagine that the backreaction of this energy generates a geometry completely different from the traditional hole with horizon. An infalling object could thus scatter off the `brick wall'  which appears where the horizon would have been, and we  get a self-consistent picture where infalling matter scatters to the outgoing radiation by a unitary S-matrix.\footnote{The form of the S-matrix was discussed in \cite{thooft1,vv}. A recent paper \cite{englert} gives a nice formulation of the 't Hooft approach, putting it on a more explicit mathematical footing.}

The problem with this approach is of course that there is no reason to take a particular `out' state of the $\{ b\}$ quanta if we want to find the effect on infalling matter. The evolution of the entire state is given by a Schrodinger equation on a complete Caucy surface (in the `nice slicing'), and individual `in-out' amplitudes give only part of the information in the wavefunctional. By the time we sum over all the relevant `in-out' amplitudes, we should recover an evolution (in the traditional general relativity setting) where the infalling object sees smooth infall through the horizon.

But with our fuzzball construction of string microstates the situation is different. Individual fuzzballs can be seen as providing a `brick wall' through which we cannot pass. But this time it is not the choice of `in-out' amplitudes that has made the difference; rather it is the structure of the microstate wavefunctional itself.

To summarize, many of the ideas that have been postulated to resolve the information puzzle had deep and interesting aspects to them. The stumbling block in each case was that there was nothing really at the horizon, and therefore one had to appeal to gauge artifacts (high temperatures in the Schwarzschild frame) or foundational issues in quantum mechanics (measuring `in-in' vs `in-out' amplitudes) to generate nontrivial effects at the horizon. But in string theory we get real degrees of freedom at the horizon \cite{membrane}, and then we can see that many aspects of these old ideas come to life by a new route.\footnote{It would also be interesting to understand better the ideas of \cite{jacobson} on getting Einstein's equations from gravitational thermodynamics. In this approach one uses the entropy of Rindler space, but it is not immediately clear what this entropy is counting. The relation of fig.\ref{fn12} suggests that one is counting fuzzball states of the gravitational theory.}

\section{Discussion}

The black hole information puzzle has been surrounded by several interrelated confusions. First, one has to note the distinction between the information paradox and the `infall problem'. In \cite{mathurrecent} it was noted that these are in a sense {\it opposite} problems; the former requires us to find some way of getting {\it different} radiation states from different microstates, while the latter asks if the infall of heavy ($E\gg kT$) objects can look essentially the {\it same} in generic microstates. 

Second, we have issue of small corrections. Here the confusion is caused by the behavior of burning paper (studied in detail in \cite{page}, for example), where the information in the paper is encoded in the radiation but very hard to decipher. This may lead to a feeling that small quantum gravity effects may have delicately encoded the black hole's information in the Hawking radiation, and the leading order Hawking computation was too crude to notice this detail. As against this is the fact there no one had produced a toy model of small corrections which {\it would} encode the information in the radiation. This issue is clarified with the inequality proved in \cite{mathurfuzz}, where it was shown that small corrections {\it cannot} encode the information and  to a nonentangled state for the radiation -- if the corrections are bounded by $\epsilon$ then the fractional decrease of entanglement due to the corrections is bounded by $2\epsilon$. This inequality was illustrated in  \cite{mathurrecent} by a simple model of small corrections which could be solved analytically. In the present paper we have used numerical computations, which allowed us to make arbitrarily  complicated models of small corrections. The monotonic growth of entanglement entropy during the evolution again illustrated the general inequality. A model of `burning paper' showed how the radiation process is unitary in this case, and fundamentally  different from the black hole radiation process. 

The inequality of \cite{mathurfuzz} has some immediate consequences. It requires that if information is to come out in Hawking radiation then there must be {\it order unity} corrections to the evolution of low energy modes at the horizon. The fuzzball construction shows that in string theory microstates are indeed very different from the traditional hole at the horizon, and for simple fuzzballs the radiation has can be seen explicitly as a unitary process. In theories that do not have the degrees of freedom to give fuzzballs (canonically quantized gravity may be one of them) we would get information loss or remnants. Another corollary of the inequality is that we cannot simply use the idea of AdS/CFT duality to argue away the information paradox: as argued in  \cite{mathurfuzz,mathurrecent}, if the AdS-Schwarzschild black hole geometry is attained in the theory as an approximate solution, then we would be forced to information loss/remnants. As another consequence of the inequality we note that several scenarios about the information problem in the literature should probably be seen as remnant scenarios. Hawking \cite{hawkingreverse} suggested that including alternate geometries in the Euclidean path integral would give a small correction to the evolution process, and restore unitarity. But since these exponentially small corrections cannot remove entanglement from the Hawking radiation, the unitarity can be restored in any such scenario only by slow leakage from a remnant. Marolf \cite{marolf} has argued that information is present at infinity at all times since the energy spectrum of a finite system is discrete; if  this energy is measurable from infinity it would identify the state of the black hole. But at this point we have to ask if the system has fuzzballs or a  traditional hole; if the latter,  then the information captured at infinity would be describing the state of a remnant since the radiation would have escaped without information.

A third puzzle has been the fact that the black hole horizon appears large and semiclassical, so it looks paradoxical that the structure here would be modified radically to a fuzzball structure which has no such horizon. In \cite{tunnel} it was noted that the classical/quantum separation of scales in the black hole is fundamentally altered by  the enormously large entropy of the black hole: the classical action for tunneling between two fuzzball microstates is of the same order as the measure of the space of fuzzball states. A crude estimate \cite{time} shows that the wavefunction of a  collapsing shell would spread over a linear combination of fuzzball states in a time much shorter than the Hawking evaporation time. 

\begin{figure}[htbp]
\begin{center}
\includegraphics[scale=.35]{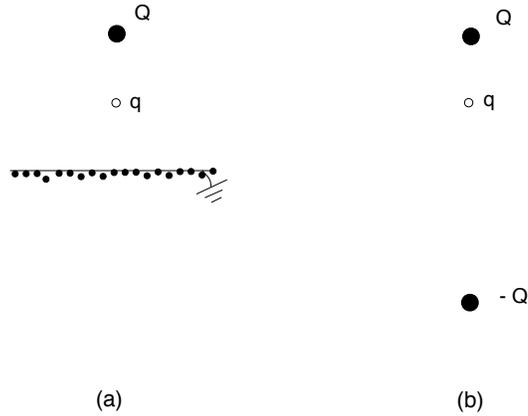}
\caption{{(a) A charge $Q$ eastablishes a field over a grounded conducting plane, and a test charge $q$ moves in this field (b) The field above the plane is reproduced by replacing the plane by an image charge $-Q$. The detailed information about the location of electrons in the conductor is lost in this replacement.}}
\label{fn10}
\end{center}
\end{figure}

Finally, there is the issue of whether the traditional black hole geometry has any role to play at all in the physics of black holes. In \cite{phase} it was noted that if we Euclidize time and perform a 1-loop path integral, then we are doing a sum over all fuzzball solutions. While each fuzzball has no symmetry, the saddle point of such a path integral, if it exists, would be expected to be the usual spherically symmetric `cigar' geometry of the Euclidean black hole. What about the Lorentzian black hole solution? This issue brings us back to the difference between the information paradox and the `infall problem'. By what was noted above about the `small corrections' issue, the information puzzle can only be solved by an explicit  construction of fuzzballs in the theory (which would   by definition modify by order unity the evolution of low energy ($E\sim kT$) quanta at the horizon.) But for the infall of high energy quanta ($E\gg kT$) we may replace a generic fuzzball by its ensemble average. At this point we used ideas in the literature (discussed in \cite{israel2,maldacena2,raamsdonk}, for example) to note that the ensemble sum over fuzzballs  may be replaced by the traditional extended black hole geometry (which does have horizons). What would be wrong would be to use this geometry with smooth horizon to compute the emission of low energy Hawking radiation; such a computation (even including small corrections) will necessarily give information loss/remnants. Instead we must recognize that individual microstates have the fuzzball structure depicted in fig.\ref{ftwop} which has a complicated quantum gravitational region  instead of a smooth horizon.  For a very crude analogy, consider the `method of images' used in electrostatics to study the field of a point charge $Q$ placed over a grounded conducting plane (fig.\ref{fn10}(a)). We can replace the conductor by an image charge $-Q$ placed at the mirror image point (fig.\ref{fn10}(b)). A test charge $q$ feels {\it almost} the same field in the region above the plane in both cases; the differences come only very close to the conductor where microscopic differences in the metal lattice determines the precise locations of electrons on the surface. The region below the conducing surface is of course very different in the two cases;  the test charge $q$ can continue smoothly past the plane boundary in fig.\ref{fn10}(b), but in this effective replacement of the conductor we have also lost the detailed information about the location of the electrons.

\b

\b

\section*{Acknowledgements}

We are grateful to Sumit Das, Steve Giddings, Stefano Giusto, Patrick Hayden, Gary Horowitz, Werner Israel, Ted Jacobson, Don Marolf,   Ashoke Sen and Edward Witten for discussions on various aspects of this problem.  We are especially grateful to Don Page for  correspondence on the issues discussed in this paper. This  work was supported in part by DOE grant DE-FG02-91ER-40690.

\end{document}